\documentclass[journal]{IEEEtran}

\usepackage{xcolor,soul,framed} 

\colorlet{shadecolor}{yellow}

\usepackage[cmex10]{amsmath}
\usepackage{array}
\usepackage{mdwmath}
\usepackage{mdwtab}
\usepackage{eqparbox}
\usepackage{url}
\usepackage{algorithmic}
\usepackage{algorithm}
\usepackage{amssymb}
\usepackage{subcaption}
\usepackage{graphicx}

\usepackage[thicklines]{cancel}

\begin{document}
\bstctlcite{IEEEexample:BSTcontrol}
    \title{Fixed-Size Dynamic Scale-Free Networks: Modeling, Stationarity, and Resilience}
  \author{Yichao~Yao,
      Minyu~Feng,~\IEEEmembership{Senior Member,~IEEE,}
      Matja\v{z} Perc,~\IEEEmembership{Member,~IEEE,}
      J\"{u}rgen Kurths\\

  \thanks{Manuscript received .  Minyu Feng and Yichao Yao was supported by the Natural Science Foundation of Chongqing (Grant CSTB2025YITP-QCRCX0007), the National Natural Science Foundation of China (NSFC) (Grant No.62206230). Matja\v{z} Perc was supported by The SlovenianResearch and Innovation Agency (Javna agencijaza znanstvenoraziskovalno in inovacijsko dejavnost Republike Slovenje) (Grant No.
  	P1-0403).}
  \thanks{Yichao Yao and Minyu Feng are with the College of Artificial
Intelligence, Southwest University, Chongqing 400715, China (e-mail:
myfeng@swu.edu.cn).}
  
  \thanks{Matja\v{z} Perc is with the Faculty of Natural Sciences and Mathematics, University of Maribor, 2000 Maribor, Slovenia, also with the Community Healthcare Center Dr. Adolf Drolc Maribor, 2000 Maribor, Slovenia, also with University College, Korea University, Seoul 02841, Republic of Korea, and also with the Department of Physics, Kyung Hee University, Seoul 02447, Republic of Korea.
}
  \thanks{J\"{u}rgen Kurths is with the Potsdam Institute for Climate Impact Research, 14437 Potsdam, Germany, and also with the Department of Physics, Humboldt University of Berlin, 10117 Berlin, Germany}
  }

\markboth{IEEE TRANSACTIONS ON SYSTEMS, MAN, AND CYBERNETICS: SYSTEMS, VOL.~, NO.~, 
}{Roberg \MakeLowercase{\textit{et al.}}: Fixed-Size Dynamic Scale-Free Networks}

\maketitle

\begin{abstract}
Many real-world scale-free networks, such as neural networks and online communication networks, consist of a fixed number of nodes but exhibit dynamic edge fluctuations. However, traditional models frequently overlook scenarios where the node count remains constant, instead prioritizing node growth. In this work, we depart from the assumptions of node number variation and preferential attachment to present an innovative model that conceptualizes node degree fluctuations as a state-dependent random walk process with stasis and variable diffusion coefficient. We show that this model yields stochastic dynamic networks with stable scale-free properties. Through comprehensive theoretical and numerical analyses, we demonstrate that the degree distribution converges to a power-law distribution, provided that the lowest degree state within the network is not an absorbing state. Furthermore, we investigate the resilience of the fraction of the largest component and the average shortest path length following deliberate attacks on the network. By using three real-world networks, we confirm that the proposed model accurately replicates actual data. The proposed model thus elucidates mechanisms by which networks, devoid of growth and preferential attachment features, can still exhibit power-law distributions and be used to simulate and study the resilience of attacked fixed-size scale-free networks.

\end{abstract}


%
\IEEEpeerreviewmaketitle


\section{Introduction}

\IEEEPARstart{M}{any} complex physical, biological, and social systems can be abstracted as networks and network science plays a crucial role in studying various complex systems \cite{barabasi2004network,wang2009cascade}. The modeling of networks that accurately reflect reality has been extensively studied for a long time. In 1998, Watts and Strogatz introduced the small-world network model \cite{watts1998collective}, imbuing network models with six degrees of separation characteristic of real-world social networks through random rewiring. Barabási and Albert later discovered the degree accumulation and preferential attachment mechanisms present in real networks, constructing network models endowed with scale-free (SF) properties \cite{barabasi1999emergence}. Subsequently, an extended SF model was proposed, allowing for edge rewiring within the network \cite{albert2000topology}. Building upon these foundational studies, various types of network models emerged. In 2001, Bianconi and Barabási noted that networks could adhere to Bose statistics and undergo Bose-Einstein condensation \cite{bianconi2001bose}. In 2003, Li et al. proposed the local-world evolutionary network model \cite{li2003local}. Later, evolutionary network models with weighted connections and community structures were established \cite{li2006modelling}, along with extensions of local-world network models incorporating node deletion considerations \cite{gu2008local}. With the research deepened into real-world networks and the diverse statistical data proliferated, the fitting of network models to the dynamic and stochastic inherent in real networks garnered increasing attention \cite{li2017fundamental,gerlach2019testing}.  The structural evolution of the network is exhibited in various important systems \cite{dagum1992dynamic,holme2012temporal}, such as neural networks \cite{sporns2004organization}, World-Wide-Web networks \cite{huberman1999growth}, etc \cite{bullmore2009complex,medo2011temporal,pi2025dynamic}. In response, evolutionary network models based on probabilistic vertex addition and deletion were constructed \cite{zhang2016random}, queueing systems were employed to describe vertex growth and removal in evolutionary networks \cite{feng2018evolving}, and various network evolution mechanisms were further investigated \cite{feng2022heritable}. Furthermore, Li et al. conceptualized nodes as queueing systems, proposing a network model with a degree birth-death mechanism \cite{li2023evolving}. Zeng et al. introduced a network model considering online and hidden vertices based on the birth and death process \cite{zeng2023temporal}. In summary, there exist a multitude of dynamic network models such as temporal networks \cite{holme2015modern}, adaptive networks \cite{raimundo2018adaptive}, and Simplicial Activity-Driven networks (SAD) \cite{petri2018simplicial} and the corresponding extents. All these models provide suitable frameworks for describing dynamic networks.

However, many of the aforementioned studies, especially those aimed at modeling SF networks, primarily focus on the modeling and evolution of networks through node growth and preferential attachment mechanisms, while paying less attention to scenarios where the number of nodes remains constant and only edges exhibit dynamics. Therefore, it is challenging to propose a model to explain the formation of SF properties in some fixed-size networks \cite{serafino2021true}, and developing a model capable of evolving SF properties within such networks is still of significant value. In many real-world systems, such as neural networks, online communication networks, and social media platforms, nodes with high connectivity often experience frequent fluctuations in connections due to their central role. For example, in a neural network, brain regions with numerous connections are more likely to form or sever connections due to fluctuating neural activity. Similarly, in email networks, central individuals frequently gain or lose contacts based on shifts in communication needs. In social media networks, popular pages tend to have highly dynamic follower bases, with significant fluctuations as users interact differently over time. These cases suggest that in fixed-size networks, the larger a node’s degree, the more frequently its connections change, underscoring the need to model degree variation as a dynamic process. To capture this characteristic, we assume that nodes with higher degrees in the network exhibit greater instability in their degree values and model the fluctuations of node degrees using a state-dependent random walk process. In our model, the probabilities of degree increase and decrease are assumed to be equal, resulting in a random but degree-dependent fluctuation process. Through the application of stochastic process theory, we analyze the stationary degree distribution under specific degree fluctuation rates. Different from the traditional SF networks, the model presented in this paper does not rely on the ``rich get richer'' principle. Rather, it is based on the principle that ``rich leads to instability''. This model achieves a dynamically stable scale-fixed network with a power-law degree distribution in scenarios where growth and preferential attachment mechanisms are not relied upon. Our proposed network model provides a framework for studying the formation and evolution of complex network systems with fixed sizes in real-world scenarios. It offers a perspective for explaining the origin of their power-law characteristics. Through this model, we can better understand, analyze, and simulate the properties and evolutionary patterns of complex networks in practice. 

The organization of this article is as follows: in Section II, we provide a detailed description of the evolutionary network model, the modeling algorithm, and a theoretical analysis of its degree distribution. The simulations regarding the degree distribution, reliability, and fidelity of the network are presented in Section III. Conclusions and future work are given in Section IV.

\section{Modeling in discrete and continuous time}
In our model, we consider a set containing all the degree values that a node can have as the state space, then regard the process of node degree variation as a state-dependent random walk within this state space. The choice of this state-dependent random walk process is motivated by its fundamental properties: unbiased increments and compatibility with Poisson-driven events. These features align well with the dynamics observed in real-world networks. Moreover,  the simplicity and adaptability of the random walk framework make analytical exploration of network degree distributions feasible, and its flexible and tunable diffusion coefficient allows the dynamic properties of the network to be adjusted. To address this random walk process, we need to consider the changes of a node's degree as an independent stochastic process. However, due to the handshake theorem, fluctuations in the degree of a node will affect not only the node itself but also its neighbors. To enable us to treat degree changes as an independent stochastic process, we do not directly alter degrees by adding or removing edges. Instead, we introduce a matching queue as an intermediate state for degree adjustment. When a node's degree increases or decreases, the node itself or its influenced neighbor will be added to the matching queue. Upon the existence of two nodes in the matching queue that can be connected, an immediate connection will be established between them, and they will get out of the queue. 

The matching queue adopts a first-in, first-out strategy to ensure that the number of edges connected to a node aligns as closely as possible with its connection demands. This design weakens the impact of a node's edge addition or removal on other nodes in the network, thereby enhancing the independence of each node's degree evolution as a stochastic process. Under this mechanism, a node's degree is influenced by others only in one specific situation: when a node has an edge demand but cannot immediately establish a connection with any other node. In such cases, its degree cannot change instantly. However, as the network evolution progresses over an extended period, this node will eventually meet a compatible node in the matching queue. Due to the queue's priority matching principle, it will connect to this node as soon as possible. Therefore, while some nodes' degree change may experience a delay, this delay is unlikely to significantly affect the steady-state degree distribution of the network when the evolution time becomes sufficiently long. The precise extent of this impact can be assessed through the simulations.

This mechanism can effectively mitigate the influence of the handshake theorem, allowing the degree evolution process of each node to remain largely independent, and it is reasonable because relationships in real networks do not only disappear in pairs. For example, in communication networks, when a high-traffic router loses connection to a neighboring router due to congestion or failure, it will still seek to establish new connections with other routers to maintain traffic flow. Similarly, in power grids, a substation may disconnect from a load to balance the grid during a fluctuation, but it will continuously seek new connections with other parts of the grid to restore balance. In cloud computing, a heavily loaded server may drop some connections to optimize performance, but it will seek new tasks or clients as the load fluctuates. These systems require constant dynamic adjustment, and the introduction of a matching queue for degree adjustments helps reflect this real-world behavior of continuously forming and dissolving connections to maintain optimal network performance.

In summary, the network model proposed in this paper is built upon the following four fundamental assumptions:

1) The degree fluctuation process of nodes in the network can be viewed as a state-dependent random walk process with a variable diffusion coefficient, where the drift coefficient increases with the node's degree.

2) When a node's degree increases, or its neighbor's degree decreases, new edges are not immediately established. Instead, the node is added to a matching queue, and a new edge is formed only when the node is matched within the queue.

3) The network does not allow isolated nodes, multiple edges, or self-loops. Therefore, nodes with a degree of 1 do not lose their remaining edge, and nodes already connected in the matching queue will not be reconnected.

4) The time a node spends in the matching queue waiting for a new connection to be established does not significantly affect the overall degree distribution of the network.

In the following subsections, we divide the modeling process into two cases: continuous time and discrete time. For each case, we present the specific modeling algorithms and analyze the degree distributions accordingly.

\begin{figure}
  \begin{center}
  \includegraphics[width=3.5in]{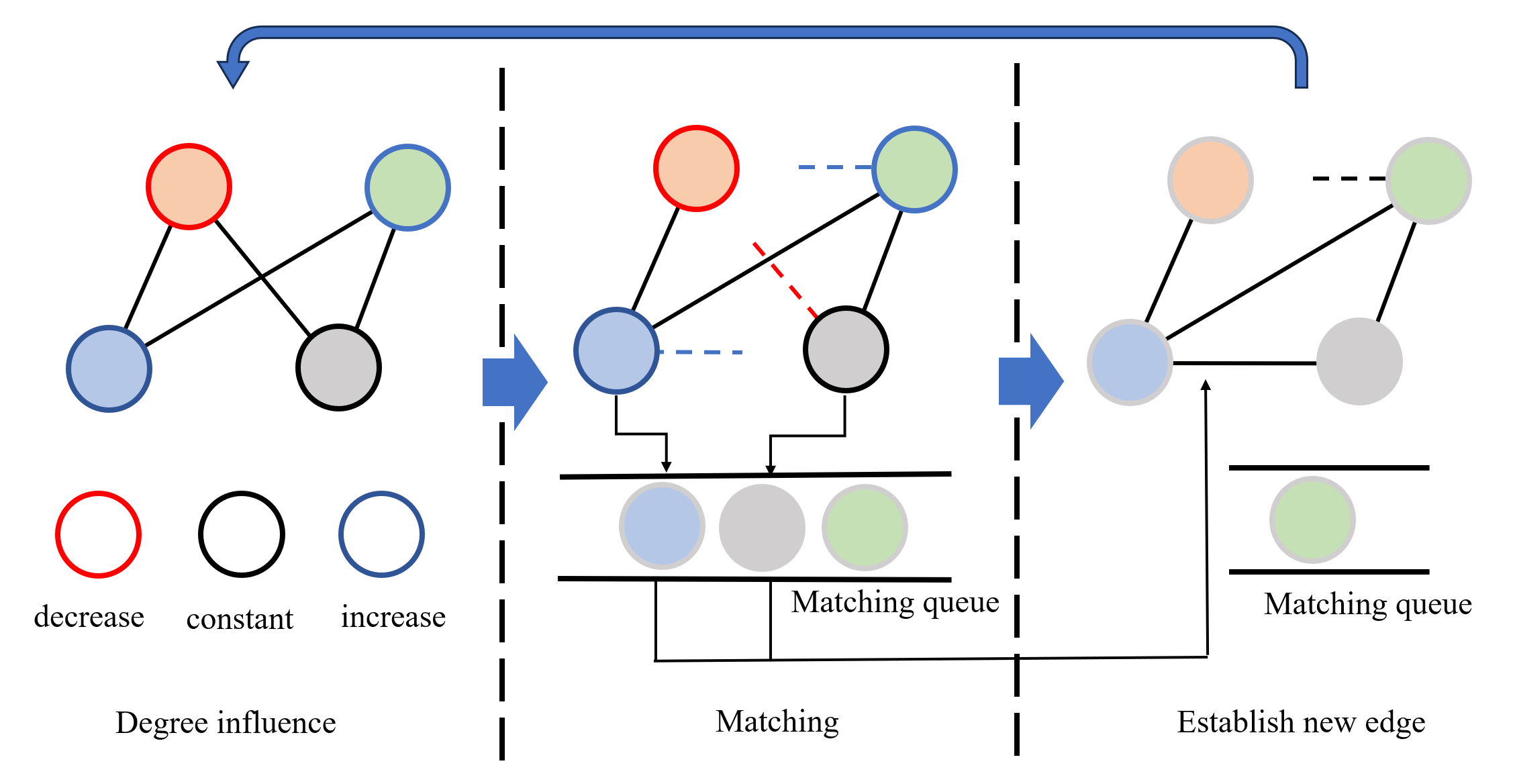}\\
 \caption{The fundamental steps of network evolution. The circles of different colors represent distinct nodes, with red, black, and blue circles indicating nodes that are about to lose a degree, maintain their degree, or increase their degree, respectively. The dashed lines represent unconnected half-edges, and nodes with half-edges will be added to the matching queue.}
 \label{f1}
  \end{center}
\end{figure}

\subsection {Discrete Time Network Modeling}

For the discrete-time case, we assign a stable probability $S(k)$ to each point in the state space, which means the probability that a node with $k$ degrees remaining at the state of having $k$ degrees after one time step is $S(k)$. That is to say, let $P$ be the one-step state transition matrix for degree changes, and let $P_{ij}$ denote the probability that a node with degree $i$ transitions to degree $j$ in one time step, then $S(k)=P_{kk}$. After that, we consider the velocity (the rate of change of the degree) of the particle's (node's) movement dependent on its position (current degree value), and fix the step size at 1. Meaning that a node's degree can only increase or decrease by 1 at each time if it changes. Furthermore, for the boundary conditions of state transition, considering that in many real-world networks, nodes with a low degree (e.g., degree $k=1$) are often subject to functional constraints. For instance, nodes cannot typically become completely isolated (degree $k=0$) due to the need for minimal connectivity in maintaining network participation or basic functionality. In addition, in the event of a failed disconnect or connection attempt, such nodes are more likely to remain in their current state rather than actively seek new connections, reflecting a conservative behavioral tendency or resource limitations. These considerations motivate us to design the transition probabilities of boundary nodes as $P_{1,1}=S(1)+\frac{1-S(1)}{2}$, and $P_{n-1,n-1}=S(n-1)+\frac{1-S(n-1)}{2}$. According to the previously discussed settings of the matching queue, the process of degree variation can be regarded as a random walk process with a diffusion coefficient that varies spatially, which possesses the Markov property. Furthermore, based on the previously mentioned assumption of ``the rich get richer'', the probability that a node chooses to change its degree in a given time step needs to be increased with its degree. Fig.~\ref{f1} presents an example of the network modeling process. As shown in the figure, nodes exhibit three basic behaviors: losing half an edge, gaining half an edge, and matching to establish a new complete edge. 

For nodes that decide to change, since we treat the degree fluctuation process as a random walk process with the step size restricted to 1, another random selection is made to either increase or decrease the node’s degree with equal probability. Furthermore, for nodes whose degree increases by 1, they are directly added to the matching queue. For nodes whose degree decreases by 1, one of their edges is randomly selected and removed, and the neighbor connected by that edge is added to the matching queue. During the evolution process, each iteration of the network begins with each node determining changes to its degree and ends once all connectable nodes in the matching queue have established connections. The complete pseudocode for the process, with the additional constraints that nodes are prohibited from forming multiple edges(duplicate edges) or self-loops, and are prevented from becoming isolated (i.e., the degree is bounded between 1 and n-1), is as the Algorithm \ref{alg:AOS1}.

\begin{algorithm}[]
    \caption{Algorithm of discrete time network model}
    \label{alg:AOS1}
    \renewcommand{\algorithmicrequire}{\textbf{Input:}}
    \renewcommand{\algorithmicensure}{\textbf{Output:}}
    
    \begin{algorithmic}[1]
        \REQUIRE T, N, G  
        \ENSURE G    

        \STATE MatchingQueue $\leftarrow$ []
        \FOR{each $t \in (1,T)$}
            \FOR{each $n \in N$}
                \STATE RandomNumber $\leftarrow$ GenerateRandomNumber(0, 1)
                \IF{$RandomNumber > \textcolor{black}{\frac{1+S(k(n))}{2}} $ and $k(n)<|N|-1$}
                    \STATE Append(MatchingQueue,$n$)
                \ELSIF{$RandomNumber \textcolor{black}{ < \frac{1-S(k(n))}{2}}$ and $k(n)>1$}
                    \STATE $m\leftarrow$ ChoiceRandomNeighbor($n$)
                    \STATE RemoveEdge($m$,$n$)
                    \STATE Append(MatchingQueue,$m$)
                \ELSE
                    \STATE CONTINUE    
                \ENDIF
            \ENDFOR
            \FOR{each $n$ in MatchingQueue}
                \FOR{each $m$ in MatchingQueue}
                    \IF{$n \neq m$ and NotHaveEdge($n$,$m$)}
                        \STATE Connect($n$,$m$)
                        \STATE Remove(MatchingQueue,$m$)
                        \STATE Remove(MatchingQueue,$n$)
                    \ENDIF
                \ENDFOR
            \ENDFOR
        \ENDFOR
            
        \RETURN G
    \end{algorithmic}
\end{algorithm}

Further, we use a Markov method to analyze the evolution of the degree distribution in the discrete-time case. The maximum degree is $n-1$, and the minimum degree is $1$, thus the state space includes $n-1$ discrete positions for nodes to reside. To keep the degree values between $1$ and $n-1$, we need to accumulate the probability of transitioning a node's degree from $1$ to $0$, and from $n-1$ to $n$, to the probabilities of remaining at $1$ and $n-1$, respectively.

\textit{Theorem 1:} If \textcolor{black}{$S(k)<1$} for all $k$, the limiting degree distribution $\Pi$ of the network exists and converges to 
\begin{align}
\left\{
\begin{aligned}
&\pi_k=\pi_1\frac{1-\textcolor{black}{S(1)}}{1-\textcolor{black}{S(k)}},1<k<n\\
&\pi_1=1/(1+\frac{1-\textcolor{black}{S(1)}}{1-\textcolor{black}{S(2)}}+\dots+\frac{1-\textcolor{black}{S(1)}}{1-\textcolor{black}{S(n-1)}})
\end{aligned}
\right.
\end{align}

\textit{Proof:} With these constraints, we can derive a transition matrix for a single node between different degrees within one time step as follows:
\begin{equation}
P=
\begin{pmatrix} 
\frac{\textcolor{black}{S(1)}+1}{2} & \frac{1-\textcolor{black}{S(1)}}{2} & 0               & \dots & 0 \\
\frac{1-\textcolor{black}{S(2)}}{2} & \textcolor{black}{S(2)}   & \frac{1-\textcolor{black}{S(2)}}{2} & \dots & 0 \\
0               & \frac{1-\textcolor{black}{S(3)}}{2} & \textcolor{black}{S(3)}   & \dots & 0 \\
\vdots & \vdots & \vdots & \ddots & \vdots \\
0 & 0 & 0 & \dots & \frac{\textcolor{black}{S(n-1)}+1}{2} \\
\end{pmatrix}          
\label{e2}
\end{equation}

According to the properties of tridiagonal matrices, when all the elements on the upper and lower diagonals of the matrix are positive, i.e. when $\textcolor{black}{S(k)}<1$ for every $k$, the corresponding Markov chain will be ergodic, and its absolute distribution will be equal to the stationary distribution. In that case, define $\pi_i$ as the proportion of nodes with degree $i$ and $\Pi=(\pi_1,\pi_2,\dots,\pi_{n-1})$ as the stationary distribution, $\Pi$ will satisfy the following equation:
\begin{equation}
    \Pi P= \Pi
\end{equation}
Calculating the stationary distribution means to solve the following system of equations:
\begin{align}
\left\{
\begin{aligned}
&(\pi_1,\pi_2,\dots,\pi_{n-1})P=(\pi_1,\pi_2,\dots,\pi_{n-1})\\
&\pi_1+\pi_2+\dots+\pi_{n-1}=1
\end{aligned}
\right.
\end{align}

When we expand the master equation formula for the degree distribution evolution, we can express it as follows:

\begin{align}
\left\{
\begin{aligned}
&\pi_1=(\pi_1(\textcolor{black}{S(1)}+1)+\pi_2(1-\textcolor{black}{S(2)}))/2\\
&\pi_2=(\pi_1(1-\textcolor{black}{S(1)})+2\pi_2\textcolor{black}{S(2)}+\pi_3(1-\textcolor{black}{S(3)}))/2\\
&\vdots \\
&\pi_{n-1}=(\pi_{n-2}(1-\textcolor{black}{S(n-2)})+\pi_{n-1}(\textcolor{black}{S(n-1)}+1))/2\\
&\pi_1+\pi_2+\dots+\pi_{n-1}=1
\end{aligned}
\right.
\end{align}

From the first and second lines of the above system of equations, we can derive:
\begin{align}
\left\{
\begin{aligned}
&\pi_2=\pi_1\frac{1-\textcolor{black}{S(1)}}{1-\textcolor{black}{S(2)}}\\
&\pi_3=\pi_2\frac{1-\textcolor{black}{S(2)}}{1-\textcolor{black}{S(3)}}
\end{aligned}
\right.
\end{align}

By mathematical induction, we can derive the following system of equations:
\begin{align}
\left\{
\begin{aligned}
&\pi_k=\pi_1\frac{1-\textcolor{black}{S(1)}}{1-\textcolor{black}{S(k)}},1<k<n\\
&\pi_1=1/(1+\frac{1-\textcolor{black}{S(1)}}{1-\textcolor{black}{S(2)}}+\dots+\frac{1-\textcolor{black}{S(1)}}{1-\textcolor{black}{S(n-1)}})
\end{aligned}
\right.
\end{align}

Therefore, combining (1)–(7), the results follow. $\hfill\blacksquare$

It is evident that the characteristic function of the stationary distribution will primarily depend on the spatial variation of the stay probability $\textcolor{black}{S(k)}$.

Specifically, based on the concept of ``rich leads to instability", we can assume the stay probability $\textcolor{black}{S(k)}$ as:
\begin{equation}
    \textcolor{black}{S(k)}=1-({\frac{k+c}{n}})^a
\end{equation}
In this scenario, nodes with larger degrees have a lower probability of remaining at that degree within a time step, ensuring that high-degree nodes are more likely to experience changes. Here, the constant $c$ serves as a smoothing parameter.  When $c = -1$ or minor, one or a set of the minimum degree in the state space becomes an absorbing state. When $c=1$ or greater, one or a set of the maximum degrees in the state space becomes a reflective state. The parameter $a$ is used to adjust the level of nonlinearity for the variation of the stay probability against the degree. For $0<k<n$, the stationary degree distribution is given by:
\begin{equation}
    \pi_k=\pi_1(\frac{c+1}{c+k})^a
\label{eq9}
\end{equation}
It is obvious that in this case, as $c$ approaches 0, the stationary degree distribution converges to a power-law distribution.

\subsection {Continuous Time Network Modeling}

To extend our approach to the case of continuous time, we now consider the increase or decrease in degree as a single event occurring on a node. Therefore, the arrival of events on a single node can be seen as a non-homogeneous Poisson process, where the inter-arrival times between two events follow an exponential distribution \cite{katti1968handbook}, and the parameter of this exponential distribution is dependent on the current degree of this node.

Similar to the discrete time intervals, for continuous time, we also define a matching queue as an intermediate process for node degree fluctuations. Specifically, the network evolution process can be divided into the following 4 steps: 
1) At the initial time, each node generates the arrival time of the next degree fluctuation event based on the exponential distribution, in which the exponential parameter is determined by its degree. 
2) Select the node with the smallest arrival time, and with equal probability, either increase or decrease its degree by 1. When the degree increases, add the node to the matching queue. When the degree decreases, disconnect the node from a random neighbor and add the neighbor to the matching queue.
3) Connect all nodes in the matching queue that can have an edge and remove them from the matching queue.
4) Subtract the arrival time of the just-updated nodes from the arrival time of all the nodes, and generate a new arrival time of the next degree fluctuation event for the just-updated node based on its new degree. Return to step 2.

\begin{figure}
  \begin{center}
  \includegraphics[width=3.5in]{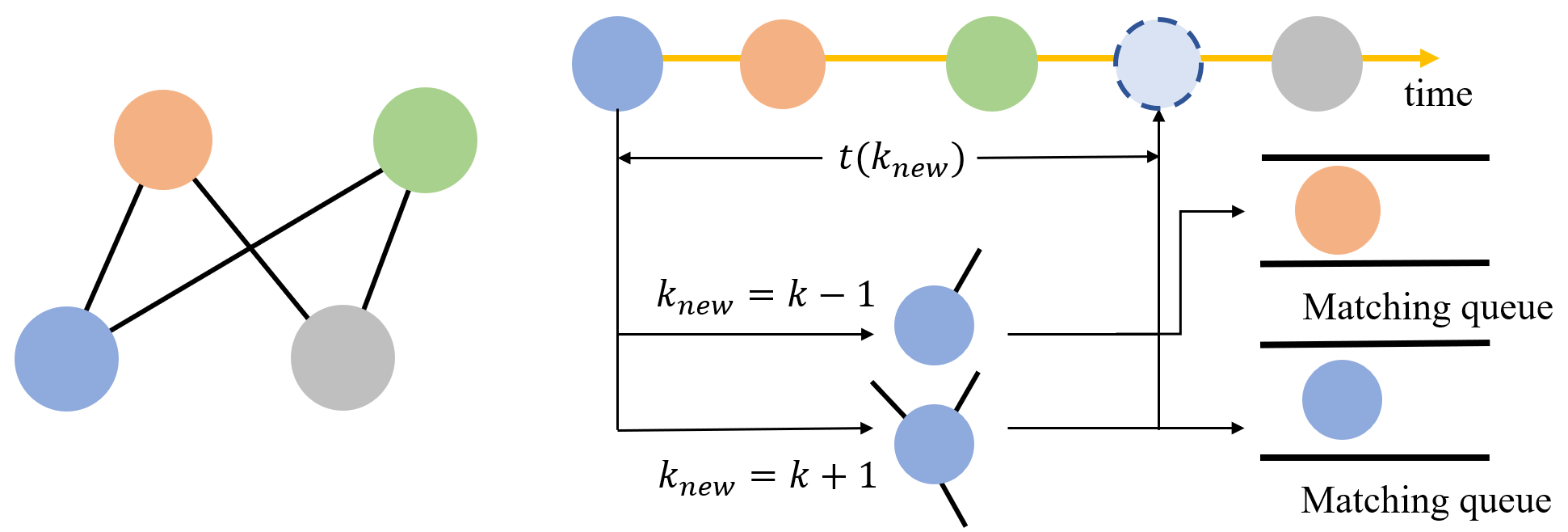}\\
 \caption{The degree fluctuation process of nodes in the continuous-time model. Nodes of different colors along the time axis represent the chronological occurrence of their degree fluctuation events. When a node's degree fluctuation event occurs, based on the status of edge disconnections and connections, either the node itself or one of its neighbors will be added to the matching queue. A round of matching will then be performed on the nodes in the queue, and based on their new degree values, the arrival times of their subsequent degree fluctuation events will be generated along the time axis.}
 \label{F2}
  \end{center}
\end{figure}

Fig.~\ref{F2} illustrates the main steps of the algorithm, and algorithm \ref{alg:AOS2} is the pseudocode for the complete modeling algorithm.

\begin{algorithm}[]
    \caption{Algorithm of continuous time network model}
    \label{alg:AOS2}
    \renewcommand{\algorithmicrequire}{\textbf{Input:}}
    \renewcommand{\algorithmicensure}{\textbf{Output:}}
    
    \begin{algorithmic}[1]
        \REQUIRE T, N, G  
        \ENSURE G    

        \STATE MatchingQueue $\leftarrow$ []
        \STATE t $\leftarrow$ []

        \FOR{each $n$ in $N$}
            \STATE \textcolor{black}{$t_n \leftarrow$ GenerateExponential($\lambda$(k(n)))}
            \STATE Append(TimeList,$t_n$)
        \ENDFOR
        \FOR{each $t \in (1,T)$}
            \STATE n $\leftarrow$ GetMinIndex(TimeList)
            
            \STATE RandomNumber $\leftarrow$ GenerateRandomNumber(0, 1)
                \IF{$RandomNumber > \frac{1}{2} $ and $k(n)<|N|-1$}
                    \STATE Append(MatchingQueue,$n$)
                \ELSIF{$k(n)>1$}
                    \STATE $m\leftarrow$ ChoiceRandomNeighbor($n$)
                    \STATE RemoveEdge($m$,$n$)
                    \STATE Append(MatchingQueue,$m$)   
                \ENDIF
                
            \FOR{each $n$ in MatchingQueue}
                \FOR{each $m$ in MatchingQueue}
                    \IF{$n \neq m$ and NotHaveEdge($n$,$m$)}
                        \STATE Connect($n$,$m$)
                        \STATE Remove(MatchingQueue,$m$)
                        \STATE Remove(MatchingQueue,$n$)
                    \ENDIF
                \ENDFOR
            \ENDFOR
            
            \FOR{each $m$ in $N$}
                \STATE TimeList(m) $\leftarrow$ TimeList(m)-TimeList(n)
            \ENDFOR
            \STATE \textcolor{black}{$t_n \leftarrow$ GenerateExponential($\lambda$(k(n)))}
            \STATE TimeList(n) $\leftarrow$ $t_n$
        \ENDFOR
            
        \RETURN G
    \end{algorithmic}
\end{algorithm}

In particular, we do not allow the emergence of multiple edges and self-loops in this work, thus, the same node and the node pairs that already have an edge in the matching queue will not be connected. Additionally, due to the boundary setting of the state space, any node whose degree reaches $0$ or $n$ will be reverted to its previous state. The continuous-time degree variation process also exhibits Markovian properties. Let $\pi_i$ be the proportion of nodes with degree $i$, $\Pi=(\pi_1,\pi_2,\dots,\pi_{n-1})$ as the stationary distribution, and $\lambda(k)$ represent the parameter of the Poisson process associated with the degree $k$. 

\textit{Theorem 2:} If $\lambda(k)>0$ for all $k$, the limiting degree distribution of the network exists and converges to a stable distribution
\begin{align}
\left\{
\begin{aligned}
&\pi_k=\frac{\pi_1\lambda(1)}{\lambda(k)},1<k<n\\
&\pi_1=1/({1+\frac{\lambda(1)}{\lambda(2)}+\dots+\frac{\lambda(1)}{\lambda(n-1)}})
\end{aligned}
\right.
\end{align}

\textit{Proof:} For state $k$ in the state space, there exists a transition rate:

\begin{align}
\left\{
\begin{aligned}
&\pi_k\lambda(k),1<k<n\\
&\frac{\pi_k\lambda(k)}{2}, k=1 or k=n-1
\end{aligned}
\right.
\end{align}
When the system is in a steady state, the rate of transitions out of each state is equal to the rate of transitions into the state. On the condition that a node can only change its degree by 1 at a time, we have the following balance condition:
\begin{align}
\left\{
\begin{aligned}
&\pi_1\lambda(1)=\pi_2\lambda(2)\\
&\pi_k\lambda(k)=\frac{\pi_{k-1}\lambda(k-1)}{2}+\\&\frac{\pi_{k+1}\lambda(k+1)}{2},1<k<n-1\\
&\pi_{n-1}\lambda(n-1)=\pi_{n-2}\lambda(n-2)
\end{aligned}
\right.
\end{align}
By mathematical induction, we get:
\begin{align}
\left\{
\begin{aligned}
&\pi_{k}=\frac{\pi_{1}\lambda(1)}{\lambda(k)},1<k<n\\
&\pi_1=1/({1+\frac{\lambda(1)}{\lambda(2)}+\dots+\frac{\lambda(1)}{\lambda(n-1)}})
\end{aligned}
\right.
\end{align}

Therefore, combining (11)–(13), the results follow. $\hfill\blacksquare$

When the formula holds, the transition rate for each degree in the state space is equal to the inflow rate. As we design the function about the Poisson parameter based on the concept of ``rich leads to instability", $\lambda$ will be increased with the degree. In this paper, we assume the $\lambda(k)$ as:  
\begin{equation}
    \lambda(k)=(\frac{k+c}{n})^a.
\end{equation}

$\lambda(k)$ is a monotonically increasing function of a node's degree $k$, ensuring that high-degree nodes are more likely to experience changes.

Furthermore, let $\pi_k(t)$ be the proportion of nodes with degree $k$ at time $t$ we can perform the analysis using the Fokker-Planck equation.

\textit{Corollary 1:} If  $\lambda(k)=(\frac{k+c}{n})^a$, the steady state distribution is a power law distribution, and the multiple steady-state solutions do not exist.

\textit{Proof:} Based on the assumptions of Poisson process, it is straightforward to derive that the variance of $\Delta k$:

\begin{equation}
    Var[\Delta k]=\lambda(k)\Delta t
\end{equation}

Thus, the diffusion term $g(k)$ in the Fokker-Planck equation is given by:

\begin{equation}
    g(k)^2=\lambda(k)=(\frac{k+c}{n})^a
\end{equation}

Since the drift coefficient $f(x)=0$, the Fokker-Planck equation is:

\begin{align}
   \frac{\partial \pi_k(t)}{\partial t}&=-\frac{\partial}{\partial k}[f(k)\pi_k(t)]+\frac{\partial^2}{\partial k^2}[\frac{g(k)^2}{2}\pi_k(t)]
   \notag
   \\
   &=\frac{\partial^2}{\partial k^2}[\lambda(k)\pi_k(t)]
\end{align}

Substituting in the concrete $\lambda(k)$,  would be:

\begin{equation}
   \frac{\partial \pi_k(t)}{\partial t}=\frac{\partial^2}{\partial k^2}[(\frac{k+c}{n})^a\pi_k(t)]
\end{equation}

In the steady-state condition, $\frac{\partial \pi_k(t)}{\partial t}$=0, the equation is reduced to:

\begin{equation}
    \frac{\partial}{\partial k}[(\frac{k+c}{n})^a\frac{\partial \pi_k}{\partial k}]=0
\end{equation}

Performing the integration, we obtain:

\begin{equation}
    \pi_k=\frac{C_1}{(\frac{k+c}{n})^a}+C_2
\end{equation}

This indicates that the steady-state distribution follows a power-law distribution with an exponent of $-a$. The distribution exhibits a single peak, thus, no multiple steady-state solutions exist. The results follow. $\hfill\blacksquare$ 

\textcolor{black}{
According to Theorem 2, for $0<k<n$, the degree distribution is:
\begin{equation}
    \pi_k=\pi_1(\frac{c+1}{k+c})^a
\label{eq21}
\end{equation}}
The inclusion of $c$ ensures that when $c>0$, the 1-degree state does not become an absorbing state, and the parameter $a$ is used to adjust the level of nonlinearity for the variation of the $\lambda$ against degree. The degree distribution will approach  an exponential when $c$ approaches 0.

\begin{figure*}[htbp]
	
    \begin{subfigure}[]{0.33\textwidth}
    	\hspace{-10pt}
        \includegraphics[width=\textwidth]{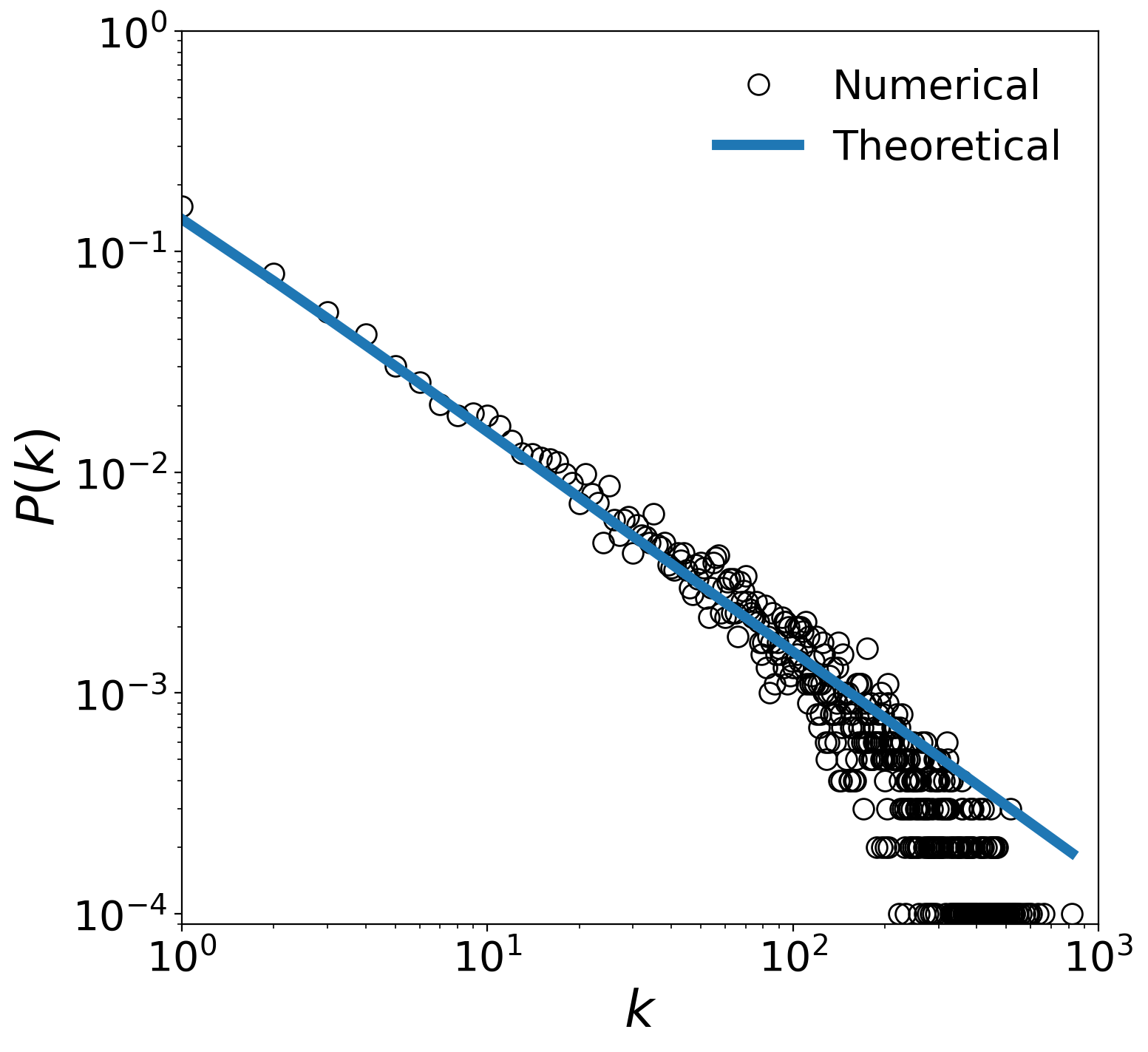}
        \caption{$a$=1}
        
    \end{subfigure}
    \begin{subfigure}[]{0.33\textwidth}
    	\hspace{-10pt}
        \includegraphics[width=\textwidth]{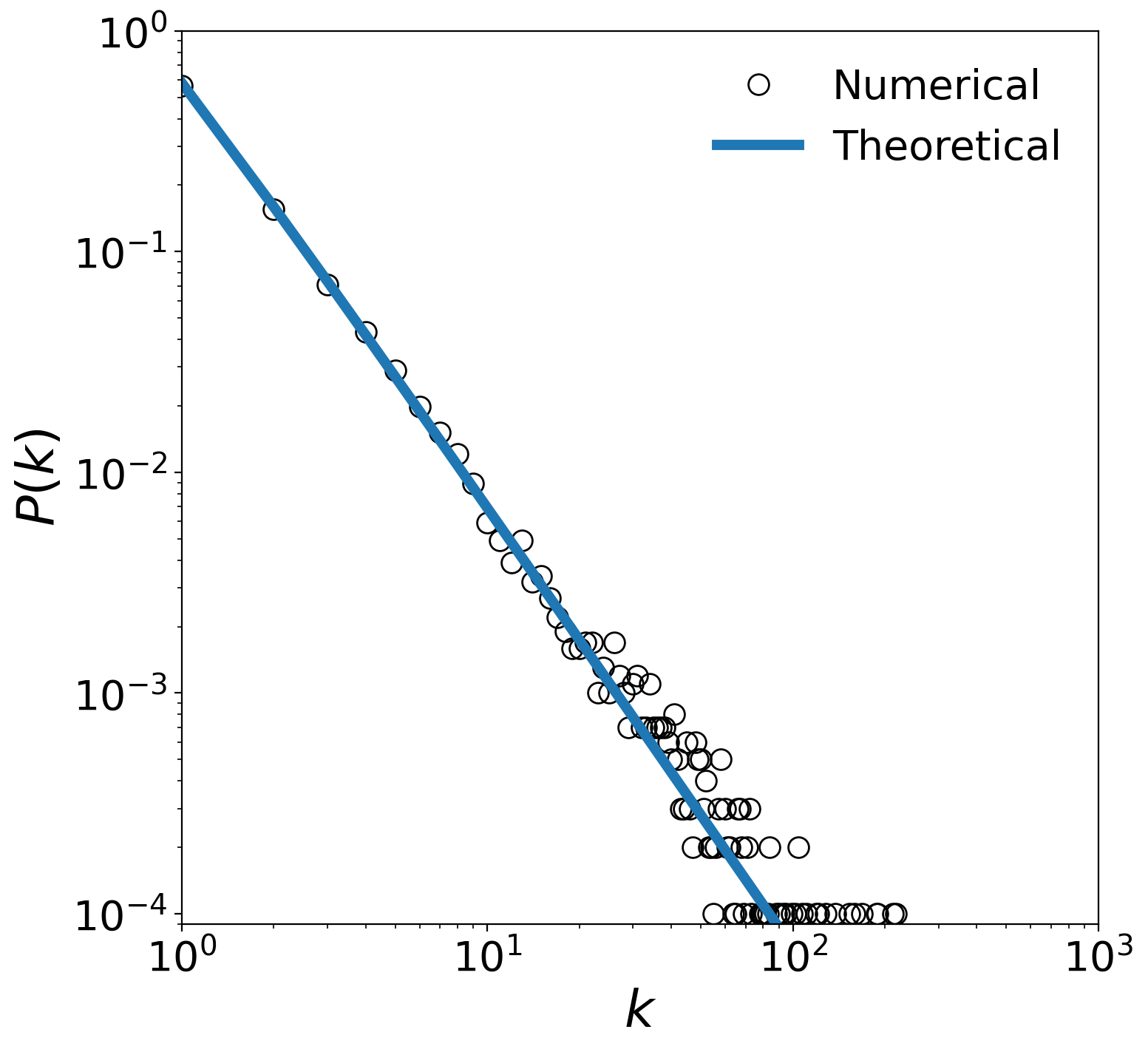}
        \caption{$a$=2}
        
    \end{subfigure}
    \begin{subfigure}[]{0.33\textwidth}
    	\hspace{-10pt}
        \includegraphics[width=\textwidth]{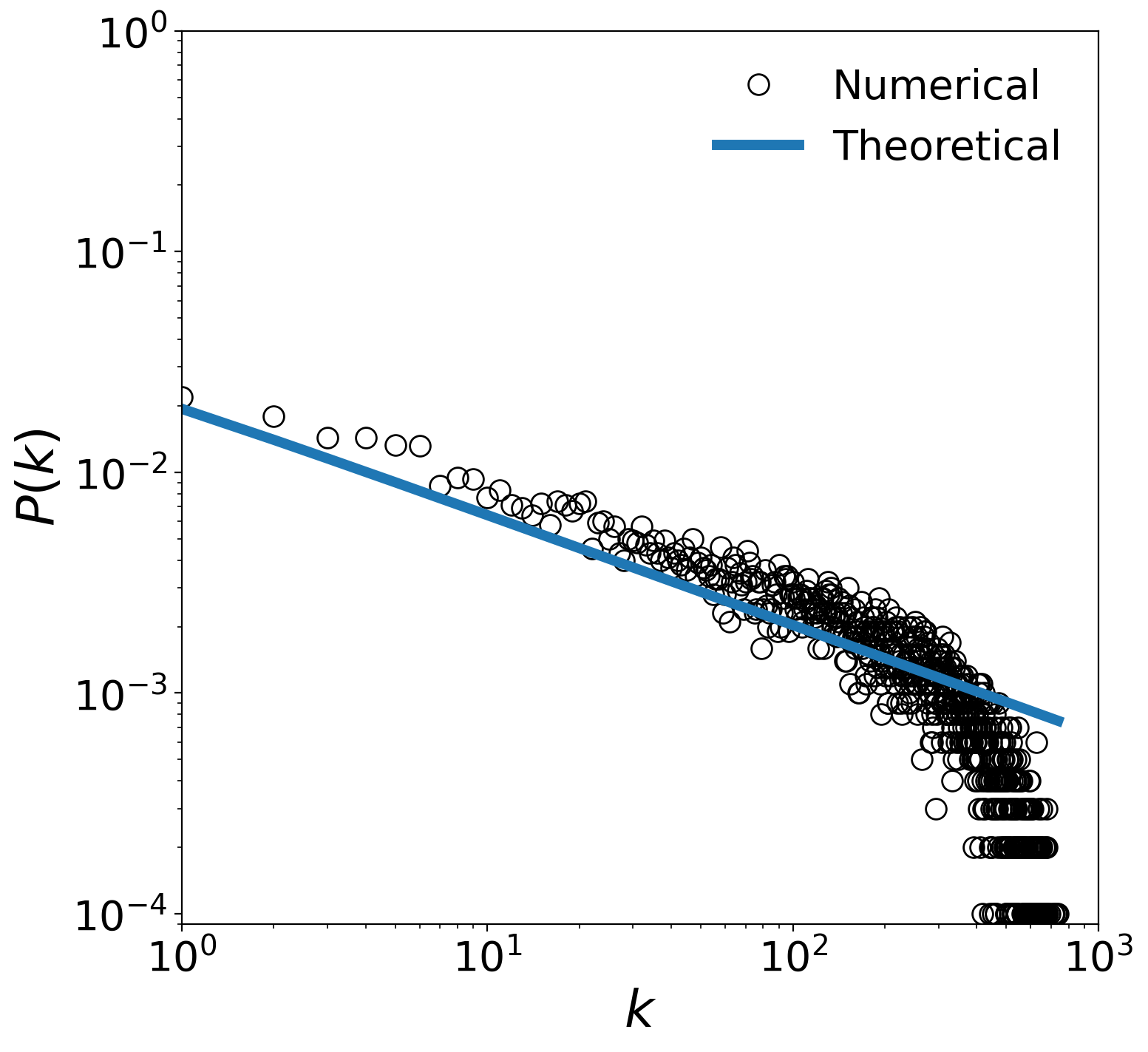}
        \caption{$a$=0.5}
        
    \end{subfigure}

\caption{\label{fig3}
The degree distribution of the network as it evolves to a stable state in discrete time. The parameter $a$ is depicted as follows: (a) $a$=1, (b) $a$=2, (c) $a$=0.5. In the figure, black circles represent the results from numerical simulations, while the blue lines represent the theoretical values.}
\end{figure*}

\begin{figure*}[!ht]
    \begin{subfigure}[]{0.33\textwidth}
    	\hspace{-10pt}
        \includegraphics[width=\textwidth]{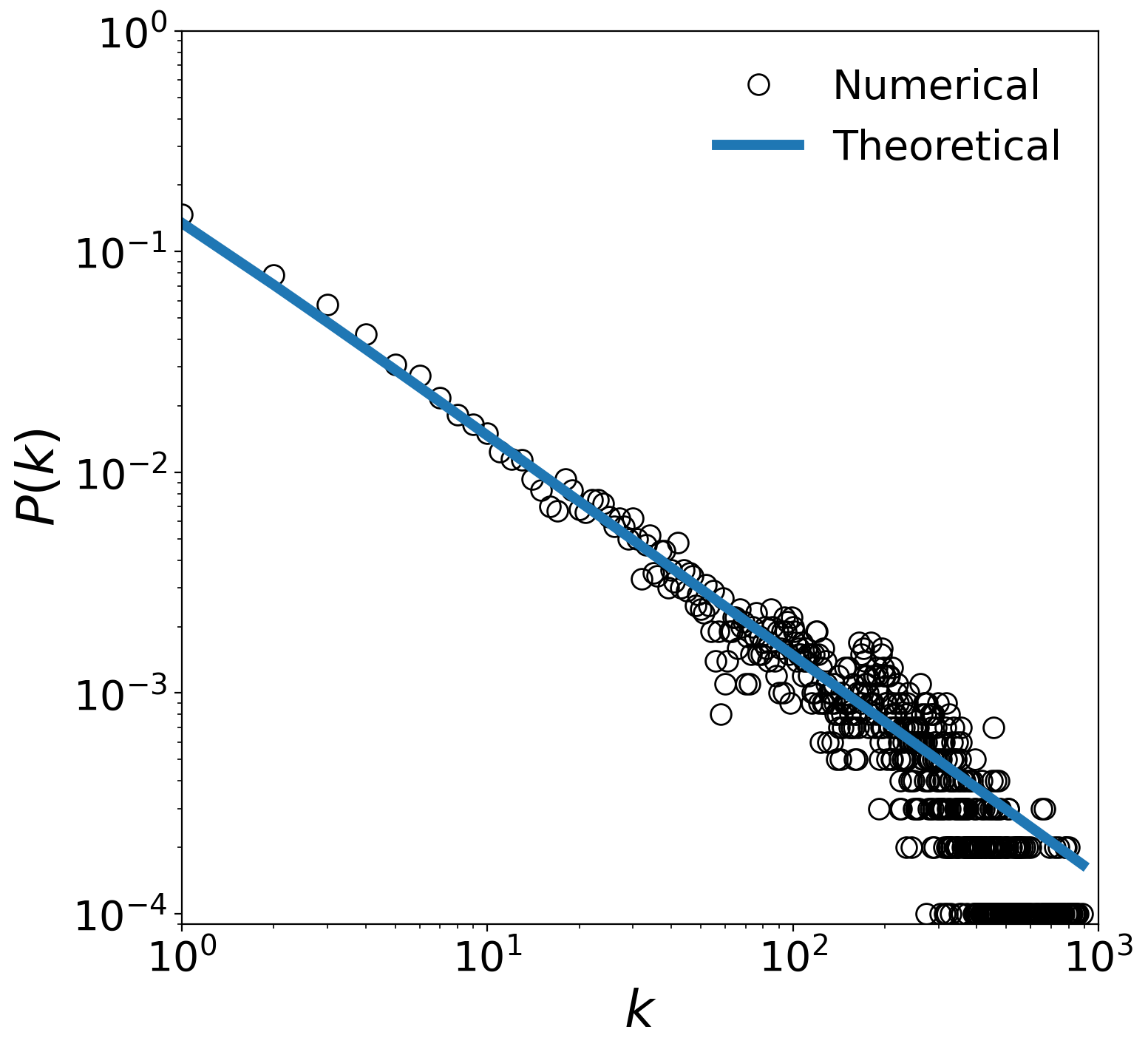}
        \caption{$a$=1}
        
    \end{subfigure}
    \begin{subfigure}[]{0.33\textwidth}
    	\hspace{-10pt}
        \includegraphics[width=\textwidth]{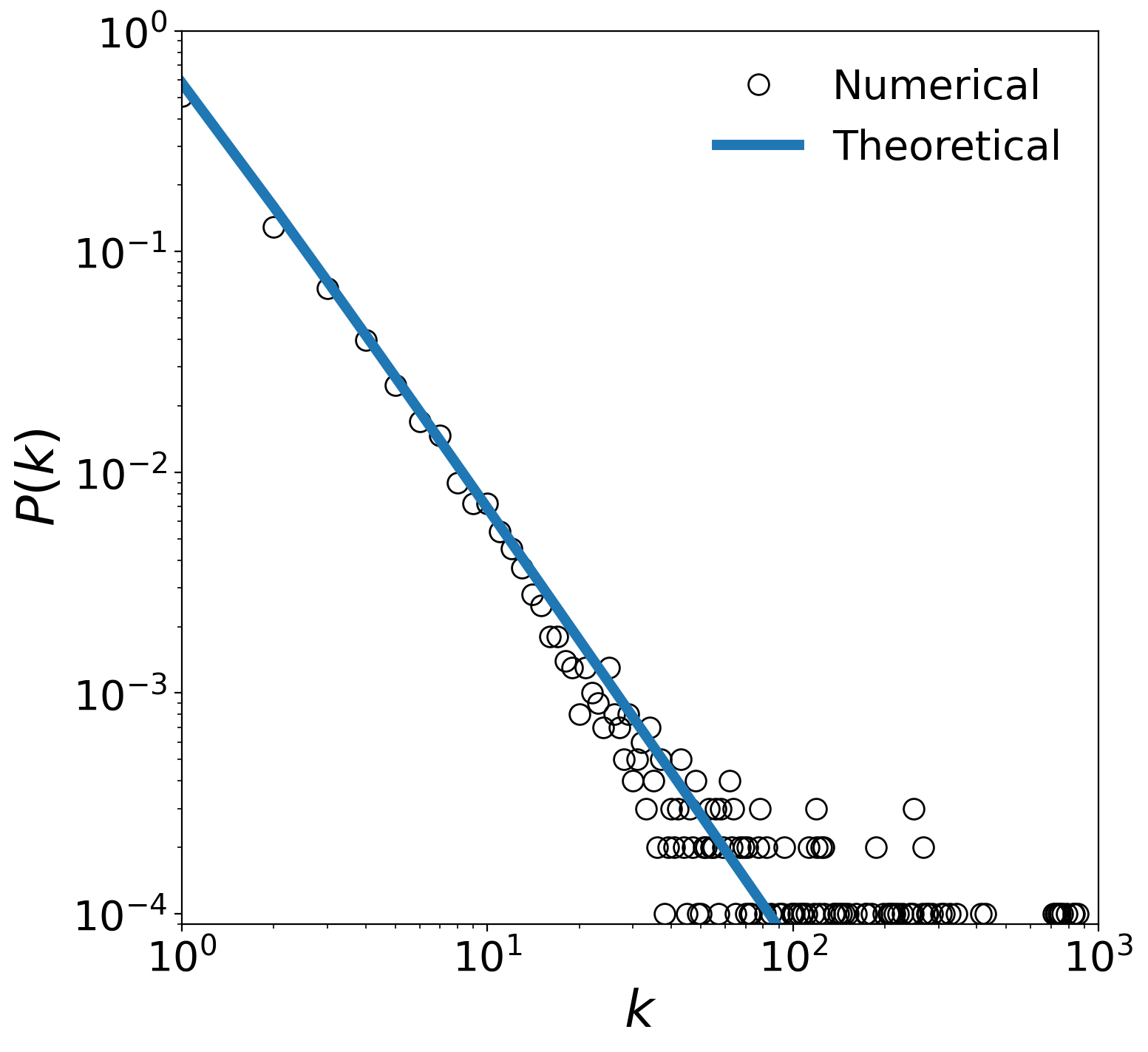}
        \caption{$a$=2}
        
    \end{subfigure}
    \begin{subfigure}[]{0.33\textwidth}
    	\hspace{-10pt}
        \includegraphics[width=\textwidth]{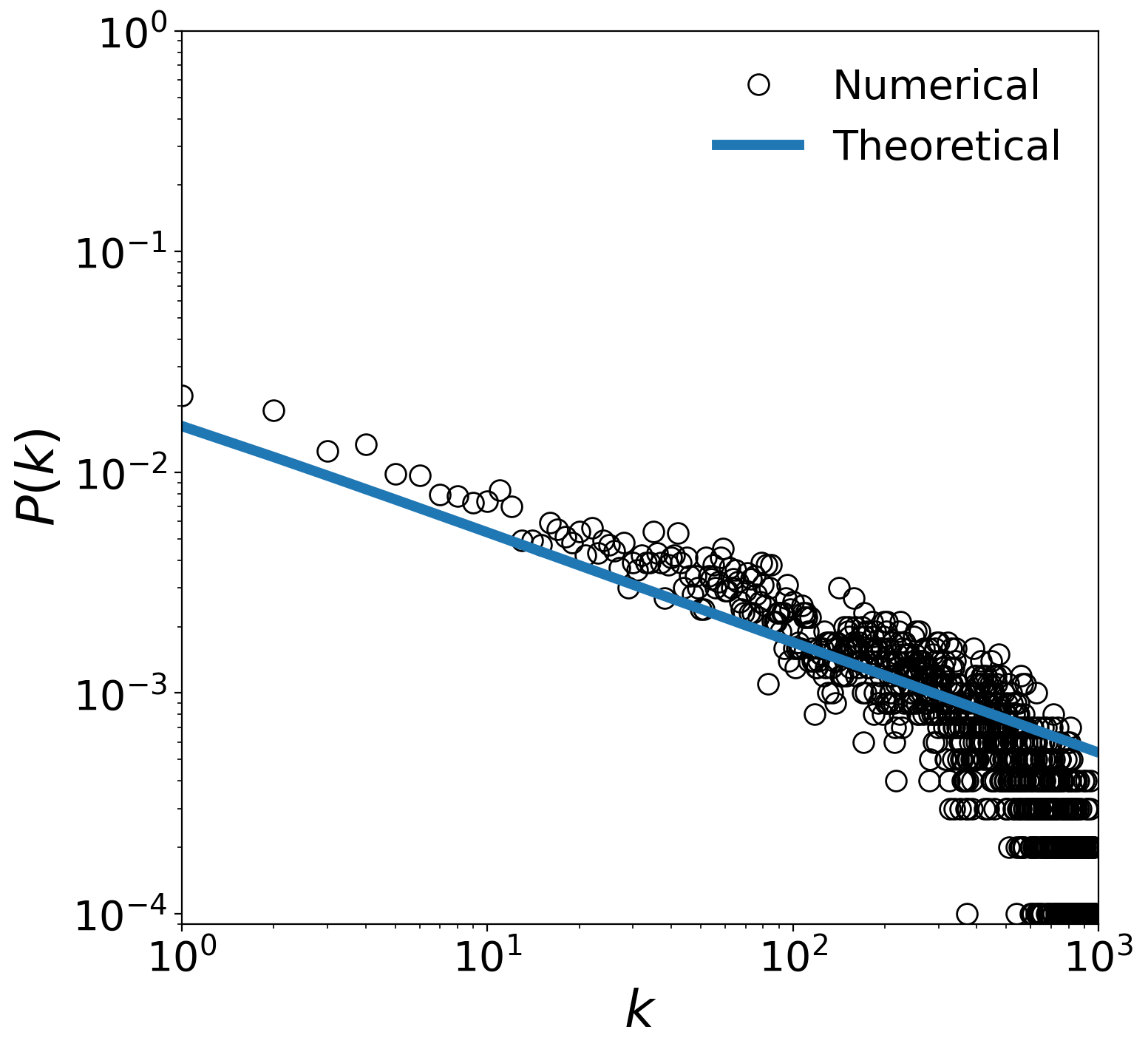}
        \caption{$a$=0.5}
        
    \end{subfigure}

\caption{\label{fig4}
The degree distribution of the network as it evolves to a stable state in continuous time. The parameter $a$ is depicted as follows: (a) $a$=1, (b) $a$=2, (c) $a$=0.5. In the figure, black circles represent the results from numerical simulations, while the blue lines represent the theoretical values.}
\end{figure*}

\section{SIMULATION}

In this section, we conduct simulation experiments to validate the previously proposed theories and models. We perform attack tests on the networks generated by the model to observe their stationarity and resilience, and use the model to fit real-world data.

In this work, all the simulations are implemented using Python 3.10. For the simulation of degree distributions and network attacks, we employed the \emph{ dense\_gnm\_random\_graph() } function from the \emph{networkx} package to generate an initial random network with a size of 1000 and an average degree of 4. During the network's evolution, the \emph{random()} function from the \emph{random} package was employed to generate random numbers to determine the probabilistic behavior of nodes at each step, while the \emph{choice()} function from the same package was used to select nodes for removal during random attacks. For fitting experiments with real networks, we used the \emph{optimize.curve\_fit()} function from the \emph{scipy} package to obtain the corresponding values of parameters $a$ and $c$ for the real networks. Subsequently, we generated initial networks of identical size to the real-world networks and an average degree of 4 for further evolution.

In the first section of this part, we simulate the network’s evolution, obtain its degree distribution, and compare it with the theoretical degree distribution. In the second section, we simulate the effects of attacks on the network and its resilience. Finally, in the third section, we apply the proposed network model to simulate real-world networks and demonstrate the fitting results for various topological properties.

\subsection{Degree Distribution and Fitting Results}

The degree distributions of the network after evolving to a steady state in discrete time are shown in Fig.~\ref{fig3}. Figs.~\ref{fig3}(a), \ref{fig3}(b), and \ref{fig3}(c) correspond to the results when $a$ is set to 1, 2, and 0.5, respectively, with $c$ fixed at 0.1. The initial network is an undirected graph with 1000 nodes and an average degree of 4. After $10^5$ iterations, the degree distributions were averaged over 10 independent runs of the evolution process to obtain the results displayed in Fig.~\ref{fig3}. It can be observed from Fig.~\ref{fig3} that the numerical simulations align well with the theoretical predictions. The lower-degree regions adhere closer to the theoretical values than the higher-degree regions. These results validate the proposed theoretical framework.

\begin{figure}[]
    \begin{subfigure}[]{0.47\textwidth}
        \includegraphics[width=\textwidth]{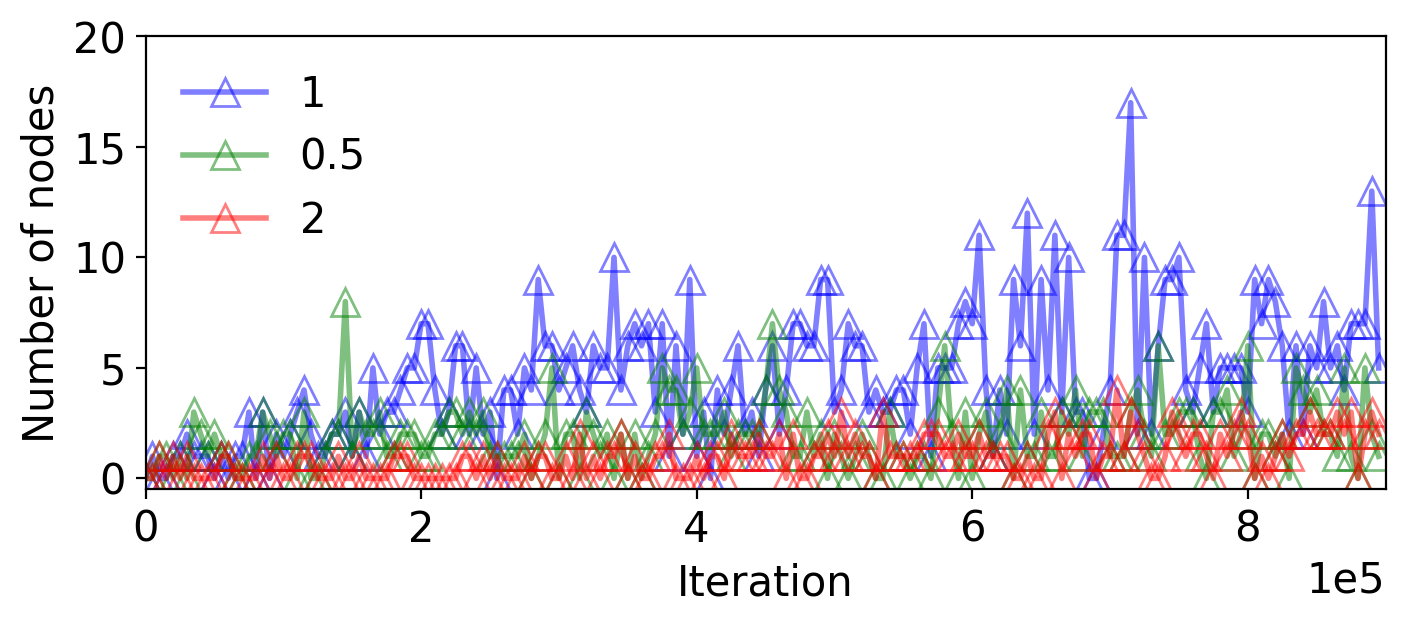}
        \caption{Discrete Time}
        
    \end{subfigure}
    \begin{subfigure}[]{0.47\textwidth}
        \includegraphics[width=\textwidth]{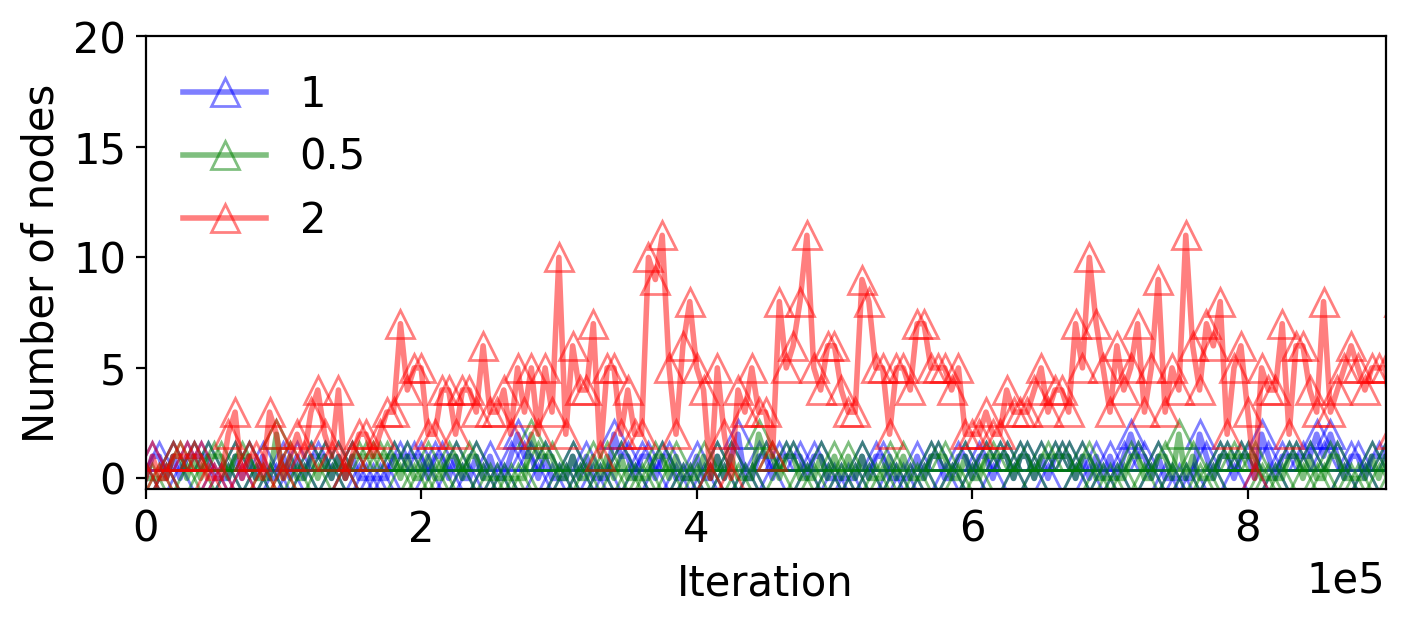}
        \caption{Continuous time}
        
    \end{subfigure}
    
\caption{\label{fig5} 
\textcolor{black}{The number of nodes in the matching queue over iterations. Subfigures (a) and (b) represent the results for discrete and continuous time, respectively. In the figures, the blue, green, and red lines correspond to the results obtained for $a=1$, $a=2$, and $a=0.5$, respectively.}
}
\end{figure}

\begin{table}[h]
\centering
\caption{\label{tab:table1}%
\centering
SIMILARITY OF THEORETICAL DISTRIBUTIONS AND SIMULATION BY KL AND JS
}

\begin{tabular}{c|cc|cc}

\textrm{}&
\multicolumn{2}{c|}{continuous}&
\multicolumn{2}{c}{discrete}\\
\hline
a         & KL     & JS    & KL    &  JS     \\
\hline
1         & 0.053  & \textcolor{black}{0.187} & 0.141 &  \textcolor{black}{0.170}     \\
\hline
2         & 0.022  & \textcolor{black}{0.200} & 0.043 &  \textcolor{black}{0.215}     \\
\hline
0.5       & 0.070  & \textcolor{black}{0.236} & 0.602 &  \textcolor{black}{0.196}     \\
\hline
\end{tabular}

\end{table}

Fig.~\ref{fig4} displays the degree distribution of networks at steady state for different values of $a$ in continuous time. Fig.~\ref{fig4}(a), Fig.~\ref{fig4}(b), and Fig.~\ref{fig4}(c) correspond to the results when $a$ is set to 1, 2, and 0.5, respectively, with $c$ fixed at 0.1. The initial network is an undirected graph with 1000 nodes and an average degree of 4. After $8\times10^7$ iterations, the degree distributions were averaged over 10 independent runs of the evolution process to obtain the results displayed in Fig.~\ref{fig4}. From Fig.~\ref{fig4}, it can be observed that the numerical simulation results closely match theoretical results, this suggests that the dynamic network model with continuous time can also achieve a good fit with theoretical values, and generate networks with stable power-law degree distributions. 

Although the theoretical results in Eqs. (9) and (21) are obtained without approximation, the network evolution involves inherent stochasticity. In the discrete-time model, nodes probabilistically increase or decrease their degrees, while in the continuous-time model, the inter-event times follow a Poisson process. Therefore, Fig.~\ref{fig3} and Fig.~\ref{fig4} not only verify the agreement between theory and simulation, but also illustrate how the stochastic fluctuations influence the degree distribution, confirming the robustness of the model.

Fig.~\ref{fig5} illustrates the variation in the number of stranded nodes in the matching queue under different parameter settings. Each point in the figure is separated by 5000 iterations. From the figure, it can be observed that in the discrete-time scenario, only when $a=1$ does the number of stranded nodes become relatively high, peaking at nearly 20. For $a=0.5$ and $a=2$, the number of stranded nodes remains consistently below 10. In the continuous-time scenario, the number of stranded nodes is higher for $a=2$, remaining under 15, whereas for $a=1$ and $a=0.5$, there are at most three nodes in the queue. Notably, throughout the whole evolution process, the number of stranded nodes in the matching queue never exceeds 2\% of the total number of nodes in the network, more than 98\% of the nodes can be considered to have independent degree evolution processes. This indicates that the theoretical analysis is largely valid, as the impact of node stranding in the matching queue on the overall network degree distribution is negligible.

To measure the similarity between the theoretical and empirical degree distributions, we utilized Kullback-Leibler (KL) divergence and Jensen-Shannon (JS) divergence \cite{zhou2020universal}. The KL divergence is used to quantify how one probability distribution diverges from a second, reference distribution. It is expressed as
\begin{equation}
    D_{KL}(\Pi||\Pi^*)=\sum_x\pi_x \log(\frac{\pi_x}{\pi^*_x})
\end{equation}
Where $\Pi$ and $\Pi^*_x$ represent the empirical and theoretical stationary degree distributions, respectively. The KL divergence is non-negative and ranges from 0 (indicating identical distributions) to $+\infty$.

\begin{figure*}[]
    \begin{subfigure}[]{0.3275\textwidth}
    	\hspace{-7pt}
        \includegraphics[width=\textwidth]{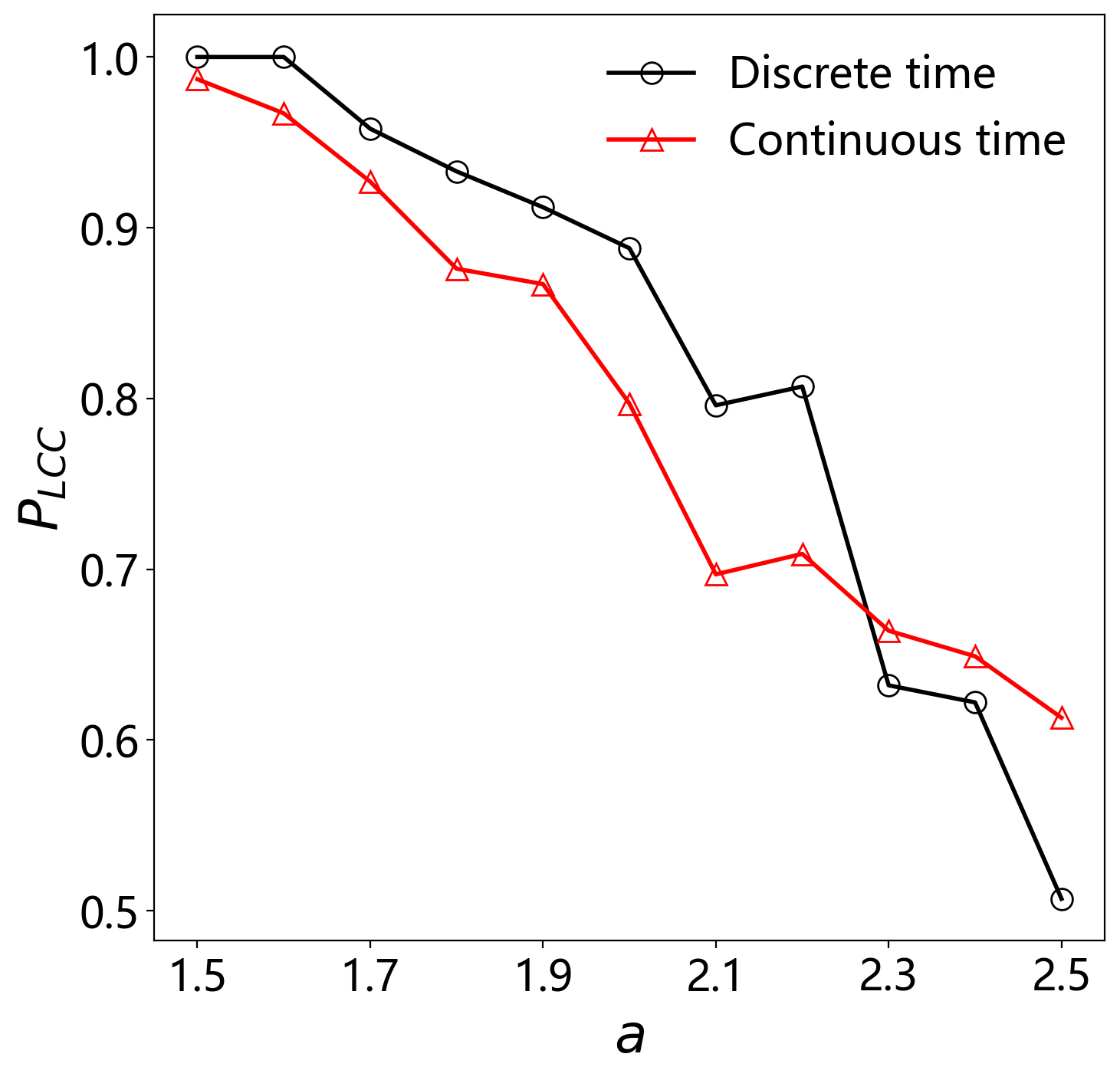}
        \caption{$P_{LCC}$ against $a$}
        
    \end{subfigure}
    \begin{subfigure}[]{0.33\textwidth}
    	\hspace{-7pt}
        \includegraphics[width=\textwidth]{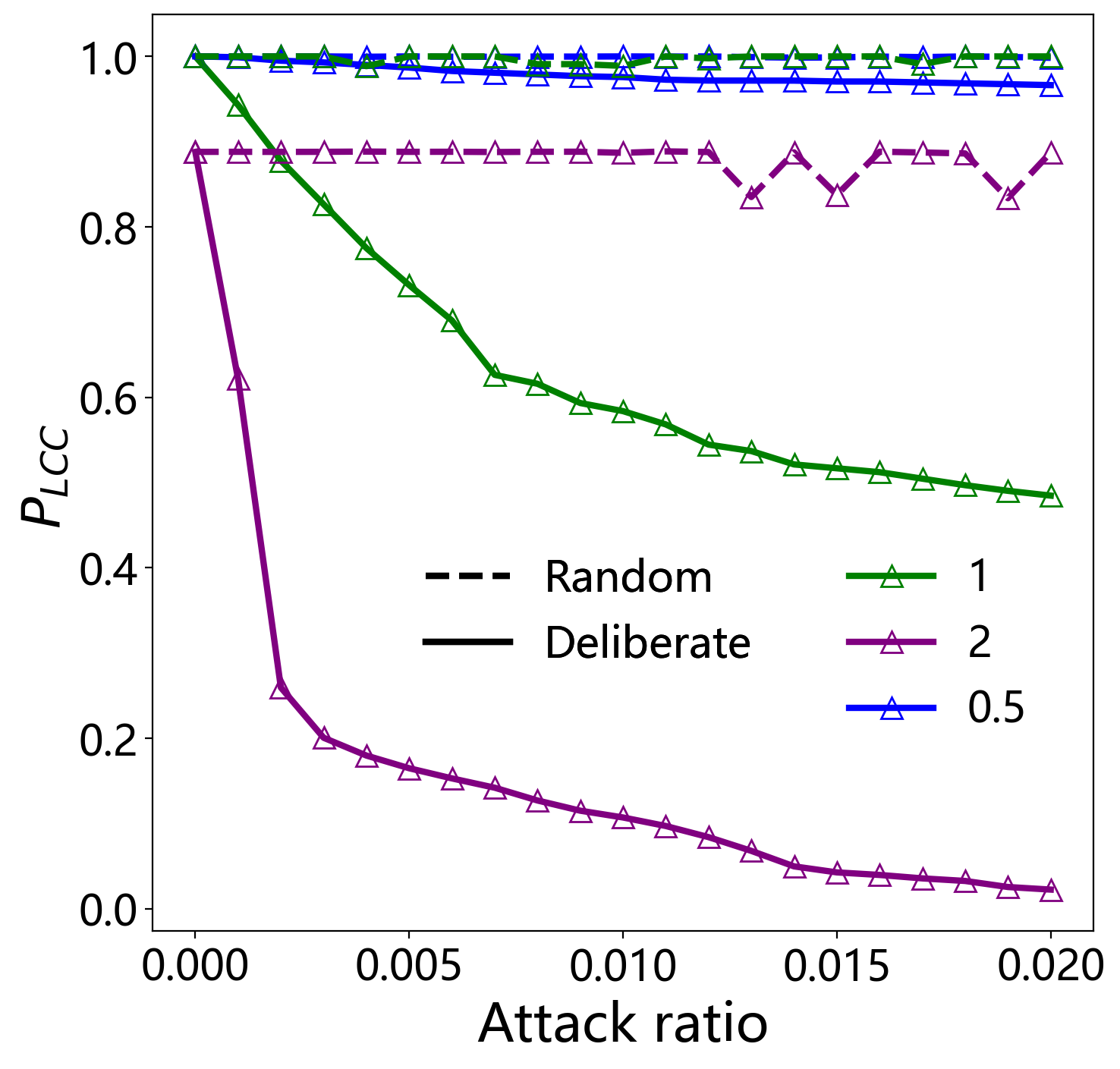}
        \caption{Continuous time}
        
    \end{subfigure}
    \begin{subfigure}[]{0.33\textwidth}
    	\hspace{-7pt}
        \includegraphics[width=\textwidth]{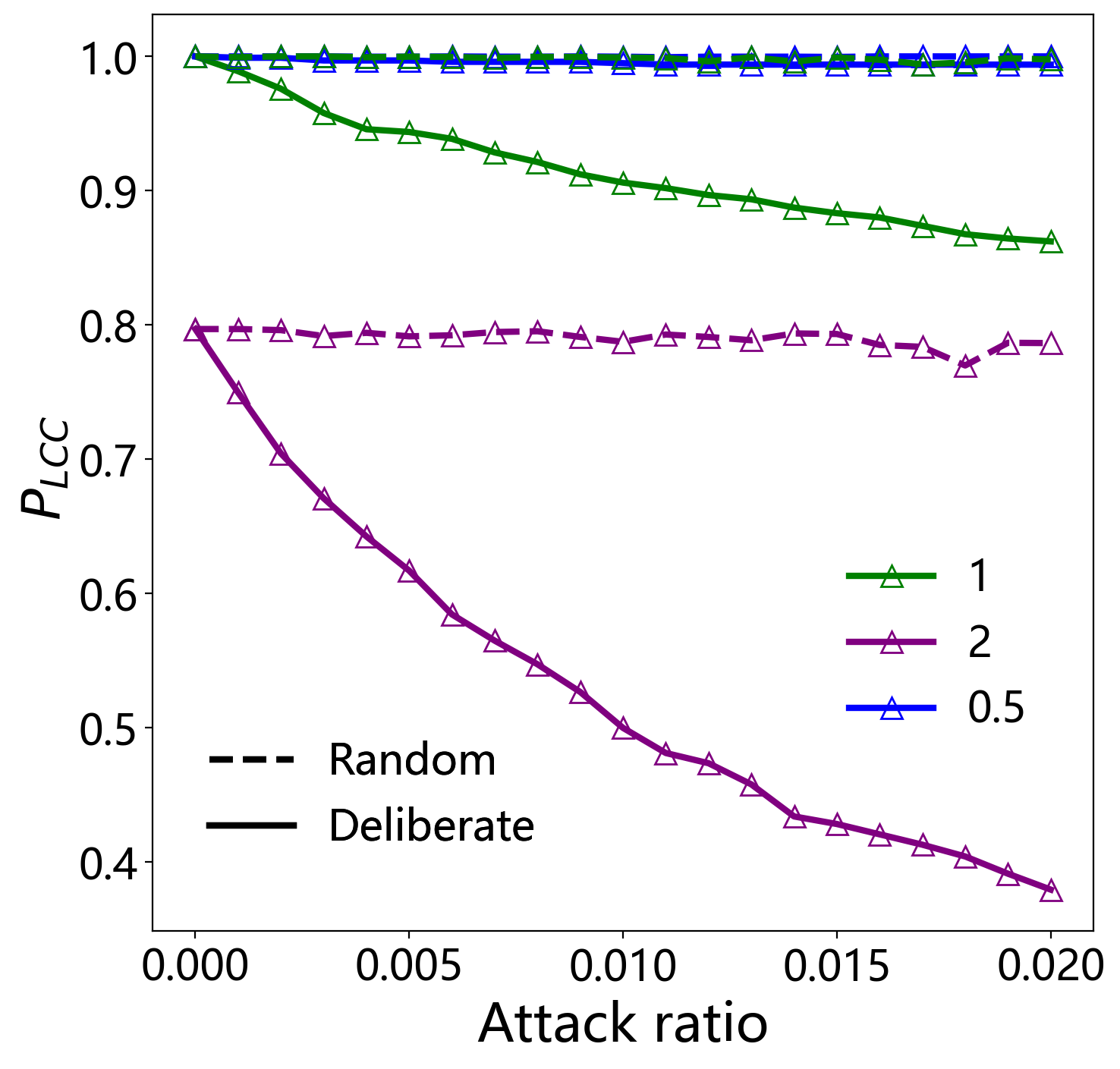}
        \caption{Discrete time}
        
    \end{subfigure}
\caption{\label{fig6} 
The impact of power-law exponent $a$ and attack ratio on the $P_{LCC}$, subplots (b) and (c) is for the model of continuous time and discrete time, respectively. In subplot (a), black and red represent results from discrete and continuous time, respectively. In subplots (b) and (c), green, purple, and blue represent networks generated by continuous time models with $a$ values of 1, 2, and 0.5, respectively, while solid and dashed lines represent deliberate attacks and random attacks, respectively.}
\end{figure*}

\begin{table*}[h]
\centering
\caption{\label{tab:table2}%
\centering
\textcolor{black}{NETWORK PROPERTIES UNDER DELIBERATE ATTACKS}
}

\begin{tabular}{c|c|c|c|c|c|c|c|c|c}

&attack&
\multicolumn{4}{c|}{continuous}&
\multicolumn{4}{c}{discrete}\\
\cline{3-10}
&ratio&$P_{LCC}$&$L$&$k_{core-max}$&modularity&$P_{LCC}$&$L$&$k_{core-max}$&modularity \\
\hline
&   0.01   & 0.583& 0.926 &   144    &  0.163 & 0.906 &   1.041  & 73& 0.075   \\
degree        &   0.02   & 0.484& 0.909 &      134 &  0.147 & 0.862 &   1.063  & 63& 0.082   \\
& 0.03 & 0.400& 0.879 & 124  &  0.132 & 0.829 &1.090 &55 & 0.089   \\
\hline
& 0.01 & 0.582 & 0.925& 144  &  0.163 & 0.907 &1.040 &73 & 0.077  \\
centrality& 0.02 & 0.484 & 0.909& 134  &  0.147 & 0.861 &1.064 &63 & 0.082  \\
& 0.03 & 0.400 & 0.87& 124  &  0.136 & 0.824 &1.089 &55 & 0.090  \\
\hline
\end{tabular}

\end{table*}

The JS divergence, which is the symmetric average of the KL divergence in both directions, is expressed as
\begin{equation}
    \textcolor{black}{D_{JS}(\Pi||\Pi^*)=\frac{1}{2}D_{KL}(\Pi||\frac{\Pi+\Pi^*}{2})+\frac{1}{2}D_{KL}(\Pi^*||\frac{\Pi+\Pi^*}{2})}
\end{equation}
Unlike KL divergence, the JS divergence is always finite and bounded within the range $[0,1]$, where 0 indicates perfect similarity and 1 indicates maximal divergence.

These two metrics provide an effective framework for measuring the differences between the theoretical degree distribution and the simulation results, as well as the discrepancies between the simulation results and real-world data. The KL divergence and JS divergence values, calculated for the theoretical degree distributions and the actual values obtained from simulations under different values of $a$, are shown in Table~\ref{tab:table1}.

From the results presented in Table~\ref{tab:table1}, we can observe that the continuous-time model consistently produces results that align more closely with the theoretical degree distribution compared to the discrete-time model. Specifically, when $a=2$, both the KL divergence and JS divergence reach their minimum values, indicating the closest match between the experimental results and theoretical values. In contrast, when $a=0.5$, both divergences reach their maximum values.

Despite these variations, the maximum KL and JS divergence values observed across all results are only 0.602 and 0.236, respectively. This indicates that the degree distribution of the network model fits the theoretical distribution well. Furthermore, as the value of $a$ increases, the simulated results exhibit an even closer fit to the theoretical values, and the continuous-time model demonstrates a stronger correspondence and better modeling performance.

\subsection{Resilience of Network Under Attacks}

Fig.~\ref {fig6}(a) illustrates the variation of the largest connected component $P_{LCC}$ with respect to the parameter $a$, where $c$ is fixed at 0.1, for both the continuous-time and discrete-time models. Fig.~\ref {fig6}(b) and Fig.~\ref {fig6}(c) present the changes in $P_{LCC}$ of the continuous-time and discrete-time models, respectively, under random and deliberate attacks of different intensities. All data points in Fig.~\ref {fig6} represent the average results obtained from experiments conducted on 10 independently generated networks with identical parameter settings. The attack ratio refers to the proportion of nodes removed from the network. Deliberate attacks sequentially remove nodes with the highest degrees, whereas random attacks remove nodes randomly. According to the results in Fig.~\ref{fig6}(a), the network is fully connected when $a<1.5$. When $a>1.5$, $P_{LCC}$ decreases as $a$ increases, showing similar trends in continuous and discrete time scenarios. Figs. \ref{fig6}(b) and (c) show that, like other SF networks, the network is robust against random attacks but vulnerable to targeted attacks, with vulnerability increasing as parameter $a$ grows. Additionally, it is observed that for the same attack frequency, the network model in discrete time demonstrates relatively stronger robustness compared to the continuous-time model.
\begin{figure}[!ht]
    \begin{subfigure}[]{0.232\textwidth}
    	\hspace{-10pt}
        \includegraphics[width=\textwidth]{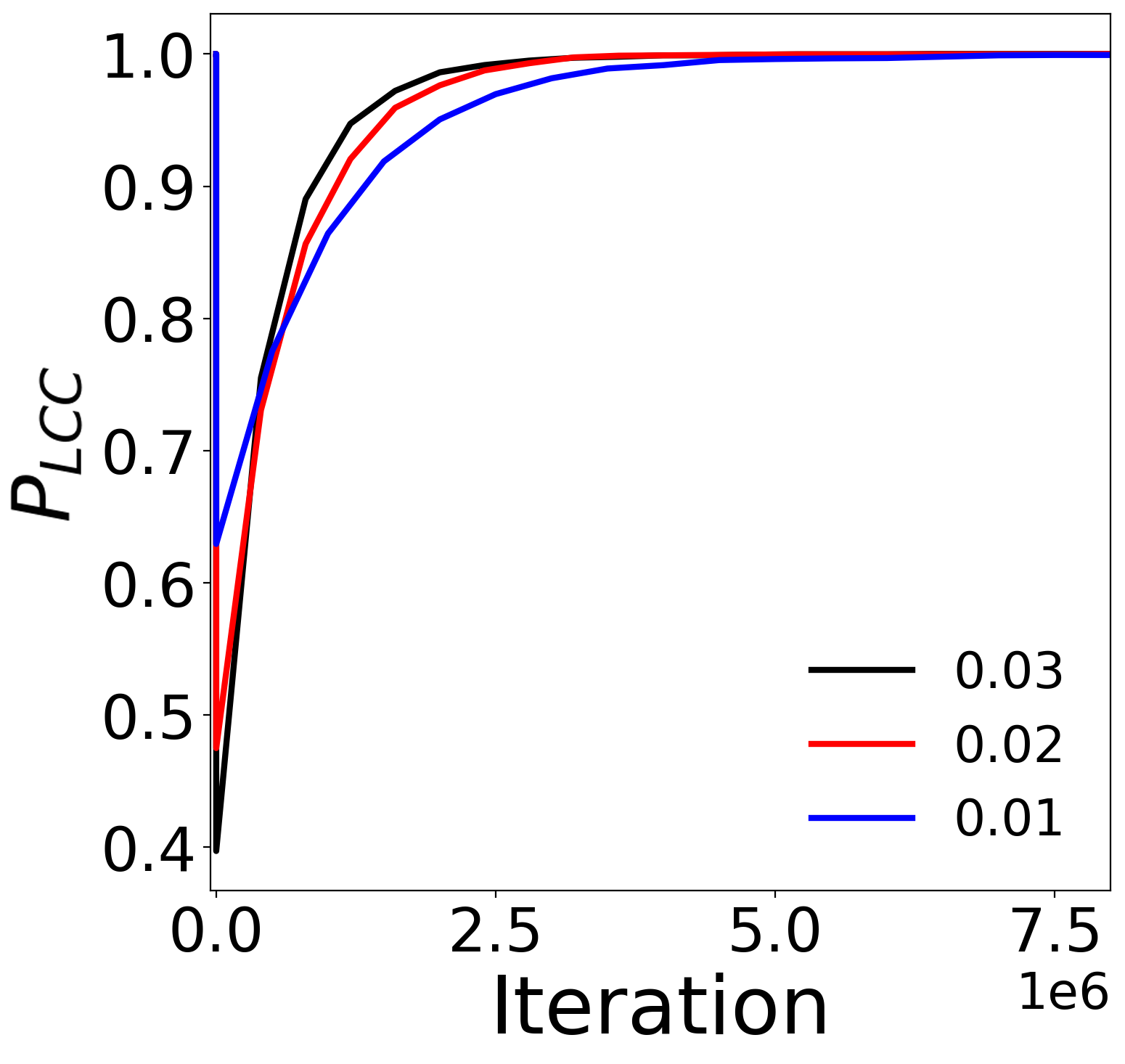}
        \caption{Recovery of $P_{LCC}$}
        
    \end{subfigure}
    \begin{subfigure}[]{0.241\textwidth}
    	\hspace{-8pt}
    	\vspace{0pt}
        \includegraphics[width=\textwidth]{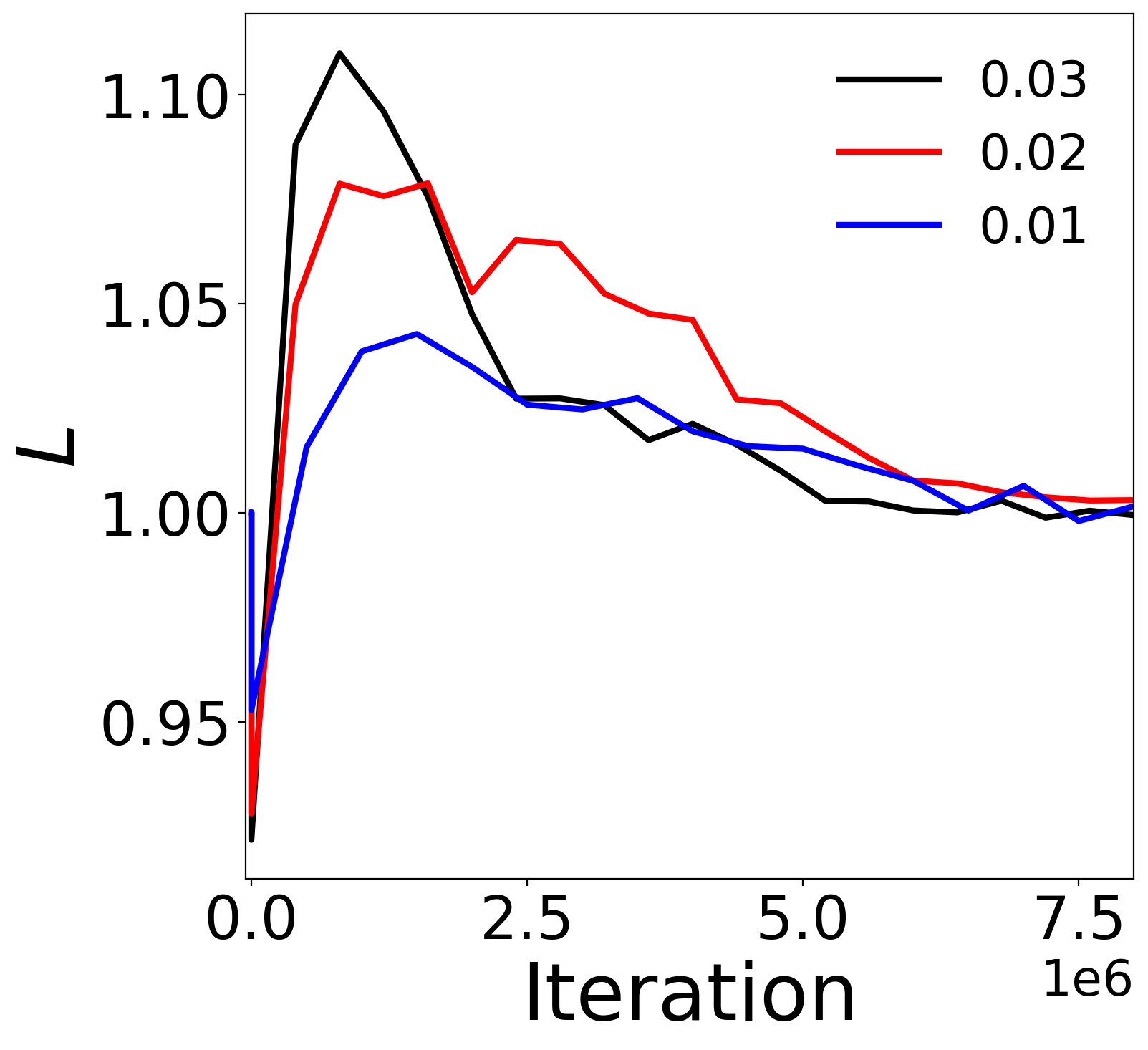}
        \caption{Recovery of $L$}
        
    \end{subfigure}
\caption{\label{fig7} 
The recovery of $P_{LCC}$ and $L$ over iterations after different ratios of deliberate attack in the continuous-time network model. $L$ represents the ratio of the current average path length to the initial average path length. The network is generated by the continuous time model with $a=1$. Blue, red, and black lines represent the results for the attack ratios of 0.01, 0.02, and 0.03, respectively.}
\end{figure}
\begin{figure}[!ht]
    \begin{subfigure}[]{0.239\textwidth}
    	\hspace{-7pt}
        \includegraphics[width=\textwidth]{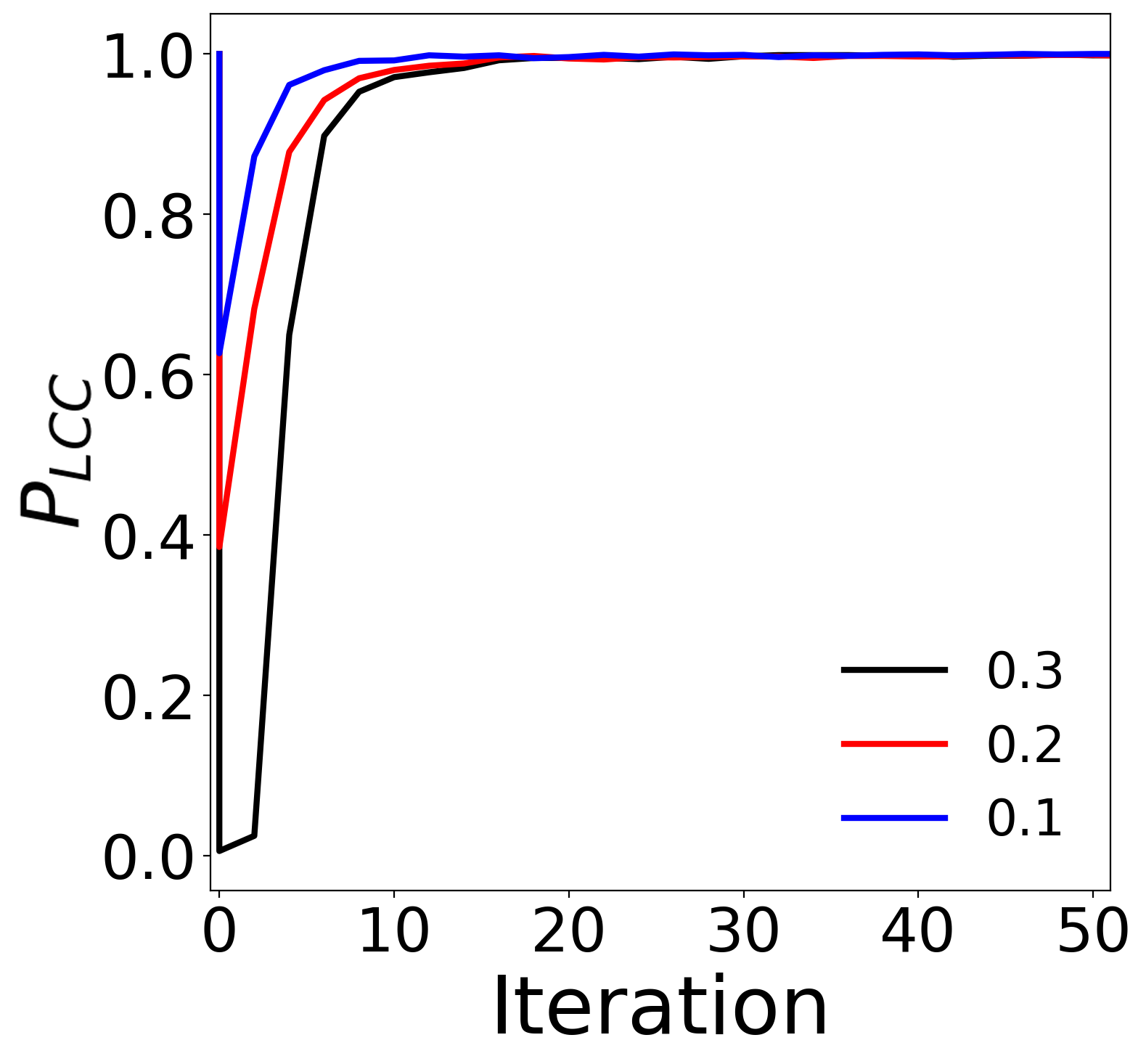}
        \caption{Recovery of $P_{LCC}$}
        
    \end{subfigure}
    \begin{subfigure}[]{0.235\textwidth}
    	\hspace{-5pt}
        \includegraphics[width=\textwidth]{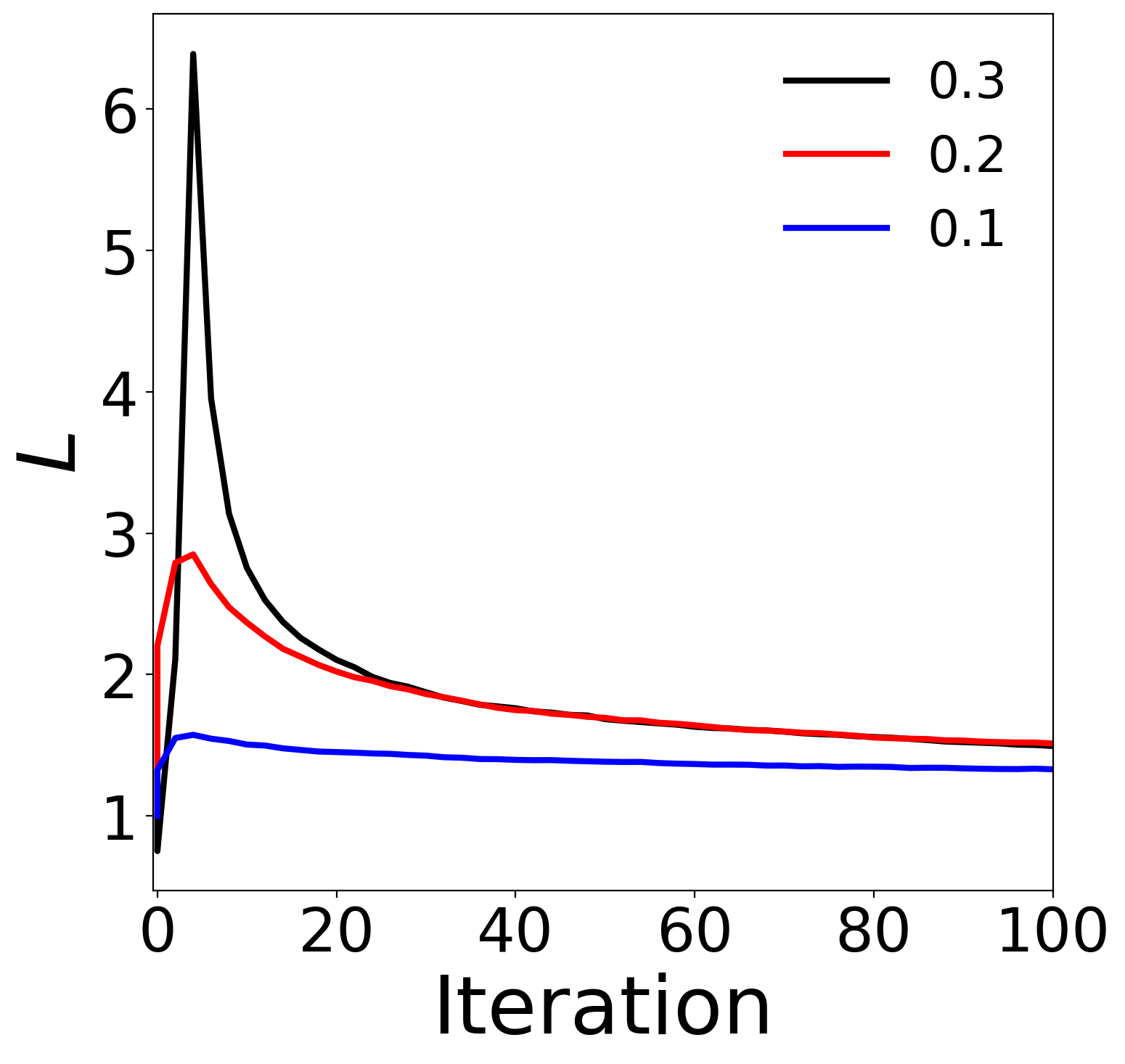}
        \caption{Recovery of $L$}
        
    \end{subfigure}
\caption{\label{fig8} 
The recovery of $P_{LCC}$ and $L$ over iterations after different ratios of deliberate attack in the discrete-time network model. $L$ represents the ratio of the current average path length to the initial average path length. The network is generated by the discrete-time model with $a=1$. Blue, red, and black lines represent the results for the attack ratios of 0.1, 0.2, and 0.3, respectively.}
\end{figure}
\begin{figure}[!ht]
    \begin{subfigure}[]{0.235\textwidth}
    	\hspace{-10pt}
        \includegraphics[width=\textwidth]{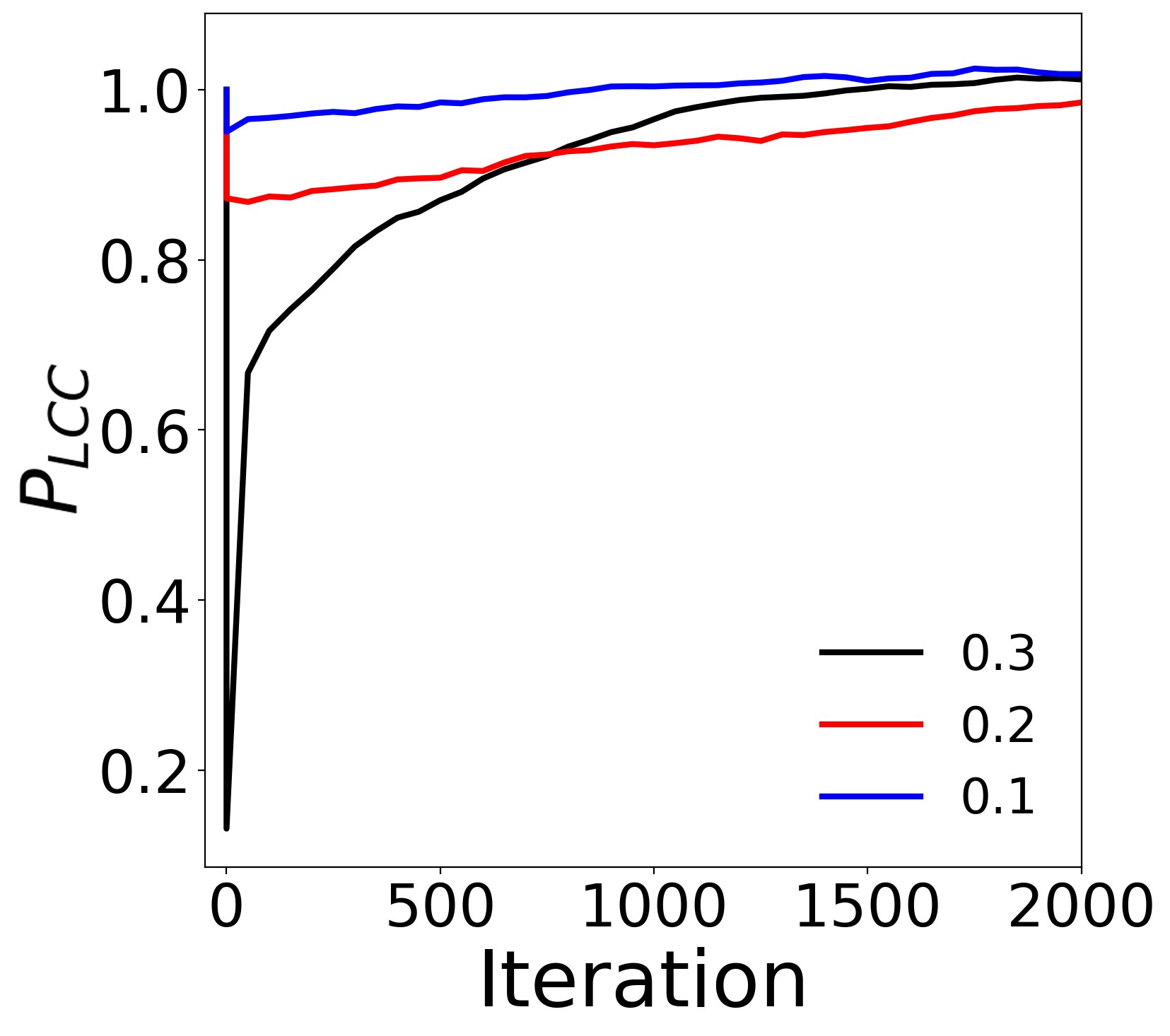}
        \caption{Recovery of $P_{LCC}$}
        
    \end{subfigure}
    \begin{subfigure}[]{0.235\textwidth}
    	\hspace{-8pt}
        \includegraphics[width=\textwidth]{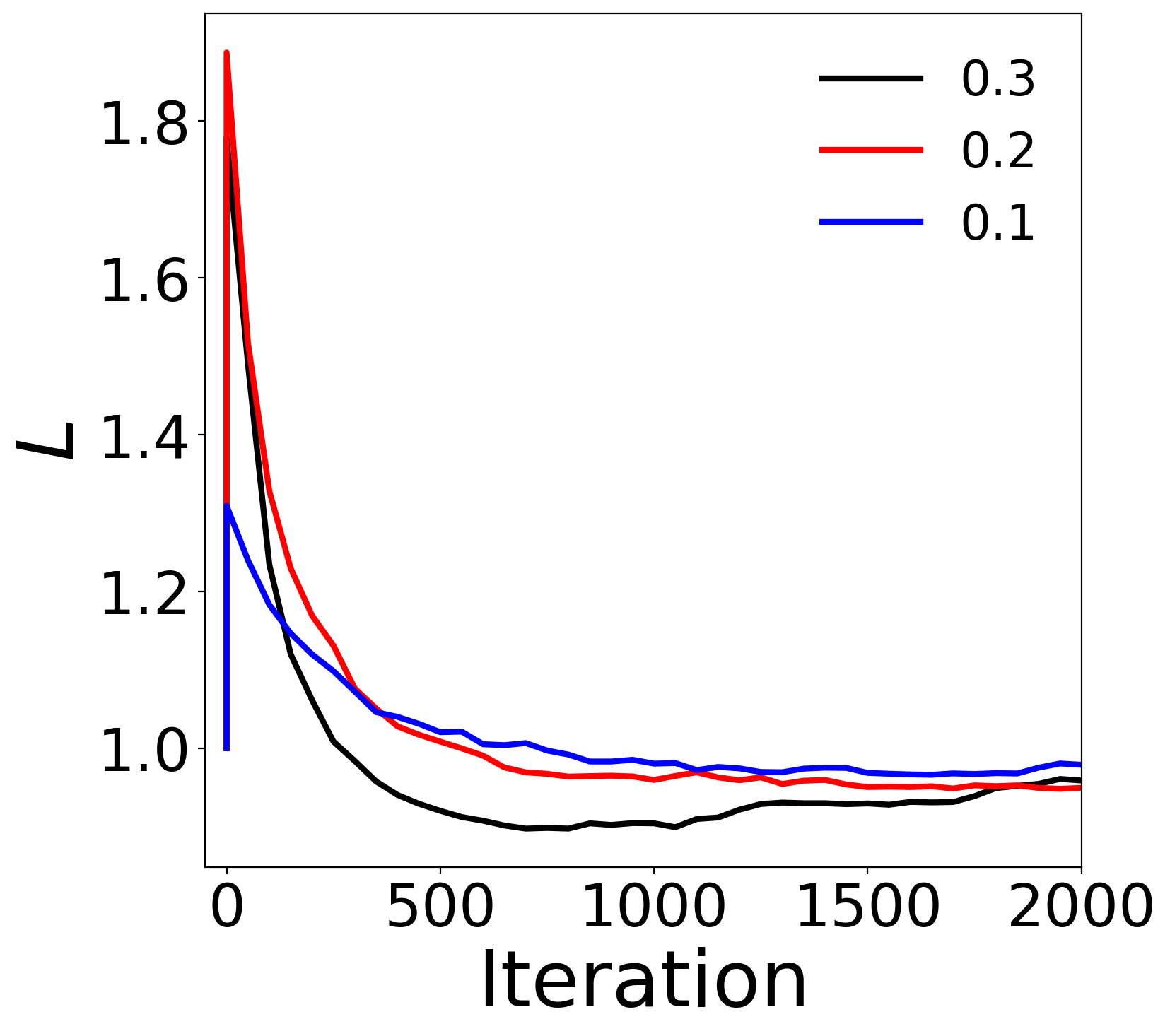}
        \caption{Recovery of $L$}
        
    \end{subfigure}
\caption{\label{fig9} 
The recovery of $P_{LCC}$ and $L$ over iterations after different ratios of deliberate attack in a dynamic network model based on a  birth-death process. $L$ represents the ratio of the current average path length to the initial average path length. The network is generated by the continuous time model with $a=1$. Blue, red, and black lines represent the results for the attack ratios of 0.1, 0.2, and 0.3, respectively.}
\end{figure}

Table~\ref{tab:table2} presents more comprehensive results. In addition to targeting nodes based on degree, it also includes results for attacks targeting nodes with the highest betweenness centrality. The table further illustrates changes in the relative shortest path length $L$, the size of the largest k-shell $k_{core-max}$, and the modularity of the network following the attacks. The $L$ is pressed as 
\begin{equation}
    L = \frac{l}{L_0}
\end{equation}
where $l$ is the average shortest path length of the current largest connected component, and $L_0$ is the average shortest path length of the network at its initial state. All data in Table~\ref{tab:table2} were obtained by conducting deliberate attacks on a network with $a=1$, 1000 nodes, and evolved to a stable state. From the data, it can be observed that the effects of attacks based on degree are nearly identical to those based on centrality. Networks generated by the discrete-time model retain a higher $P_{LCC}$ under attack, while those generated by the continuous-time model exhibit slightly higher modularity.

Furthermore, since the network continuously converges during its evolution, there is a recovery after an attack. Figs.~\ref{fig7} and \ref{fig8} display the resilience of the network following an attack. Subfigures (a) and (b) in Figs.~\ref{fig7} and \ref{fig8} respectively depict the changes in the largest connected component $P_{LCC}$ and the relative shortest path length $L$ under different deliberate attack frequencies for the continuous and discrete time models, the data points in Figs.~\ref{fig7} and \ref{fig8} represent the average results obtained from attacks performed on 10 independently generated networks with identical parameter settings. Although the recovery of $P_{LCC}$ is a natural result of the rewiring mechanism, Figs.~\ref{fig7}–\ref{fig9} are presented to illustrate the detailed recovery trajectories of $P_{LCC}$ under different parameter configurations. By comparing the recovery processes and timescales associated with various values of attack ratio, we can assess how it affects the network’s resilience. In addition, comparing the recovery patterns of $P_{LCC}$ and the relative average shortest path length $L$ provides further insight into the distinct temporal responses of global connectivity and path efficiency during the restoration process. The results shown in Figs. 7 and 8 indicate that, even though the size of our network is fixed, it can naturally recover its topological properties after being subjected to attacks, similar to some evolving networks.

For both the continuous and discrete time networks, the parameters are set to $a=1$ and $c=0.1$. The attack frequencies for the continuous-time network are set to 0.01, 0.02, and 0.03, while for the discrete-time network, they are 0.1, 0.2, and 0.3. From Figs.~\ref{fig7} and \ref{fig8}, it can be observed that $P_{LCC}$ gradually recovers after the attack, and $L$ initially increases with the growth of $P_{LCC}$. After $P_{LCC}$ recovers to its maximum value, then $L$ gradually decreases to 1 as the network evolves. The total time for the average shortest path length to fully recover is longer than that for connectivity. Additionally, a notable pattern is observed in the recovery after high-ratio attacks: although a higher attack ratio results in a greater reduction in network connectivity, the resilience is stronger in that case compared to when the attack ratio is lower. This is because a higher attack ratio creates a larger hub node gap, reducing competition pressure among other nodes, and allowing new hub nodes to appear more quickly. Additionally, we introduced a dynamic scale-free network based on a birth-death process for comparison \cite{feng2024information}, with the results presented in Fig.~\ref{fig9}. The results in Fig.~\ref{fig9} are also averaged over independent attacks on 10 independently generated networks with identical parameter settings, the attacks were deliberate, with attack rates of 0.1, 0.2, and 0.3. From the figure, it can be observed that this network demonstrates stronger attack resistance compared to our model. However, the regularity of its recovery process is less pronounced. Specifically, $L$ is most affected at an attack rate of 0.2, while the $P_{LCC}$ recovery curve at an attack rate of 0.3 differs significantly from those at attack rates of 0.2 and 0.1.

\subsection{Fitting Real Networks}

\begin{figure}[!ht]
    \begin{subfigure}[]{0.235\textwidth}
        \includegraphics[width=\textwidth]{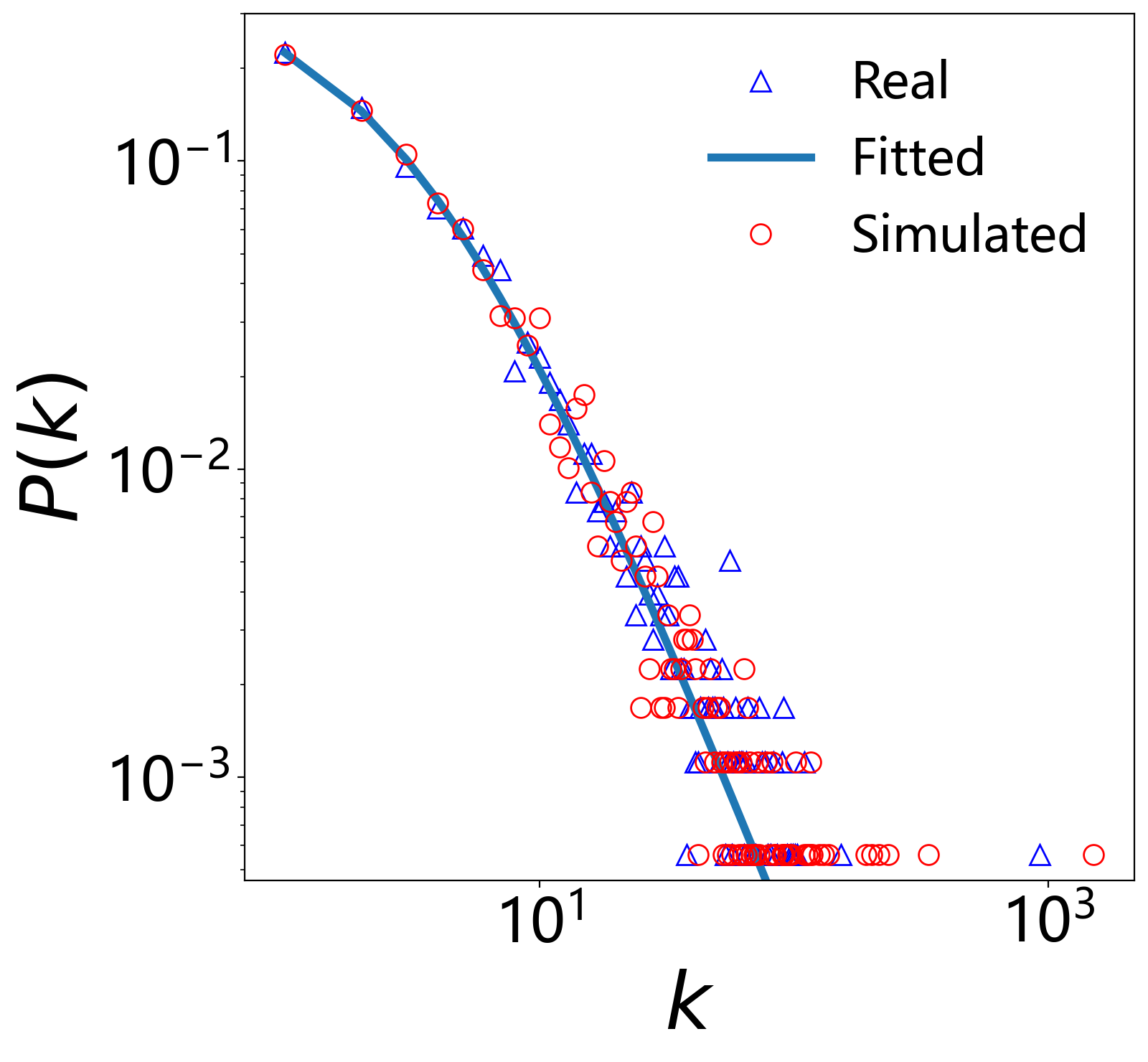}
        \caption{Continuous Time}
        \label{fig5:sub1}
    \end{subfigure}
    \begin{subfigure}[]{0.237\textwidth}
        \includegraphics[width=\textwidth]{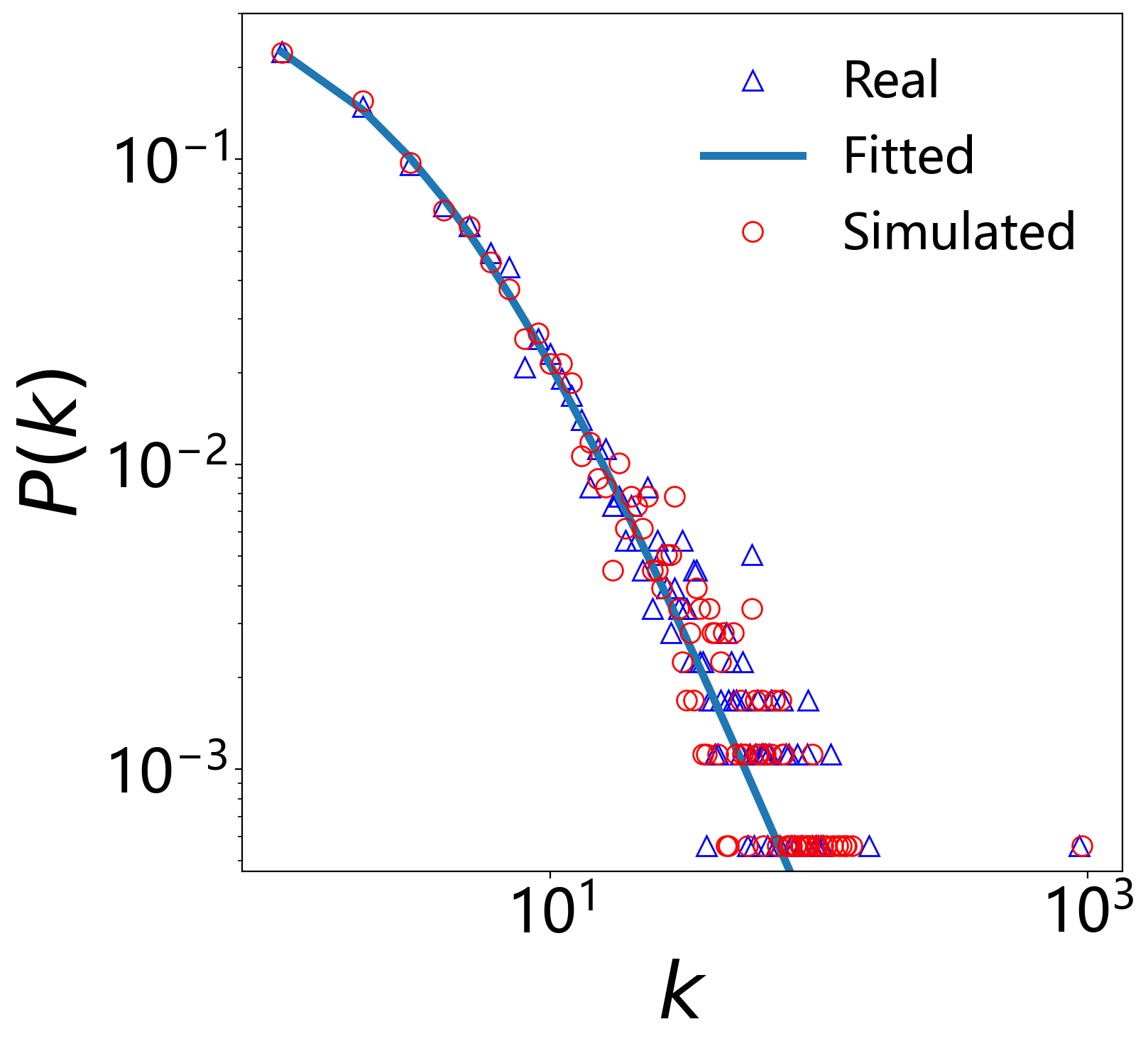}
        \caption{Discrete Time}
        \label{fig5:sub2}
    \end{subfigure}

\caption{\label{fig10}
The fitting results of a continuous and discrete time network model to a Fly Brain Network (FBN). In these subplots, the blue triangles represent data from the real networks, the blue lines represent the fitting curves, and the red circle denotes the data obtained from the simulation network. }
\end{figure}

\begin{figure}[!ht]
    \begin{subfigure}[]{0.235\textwidth}
        \includegraphics[width=\textwidth]{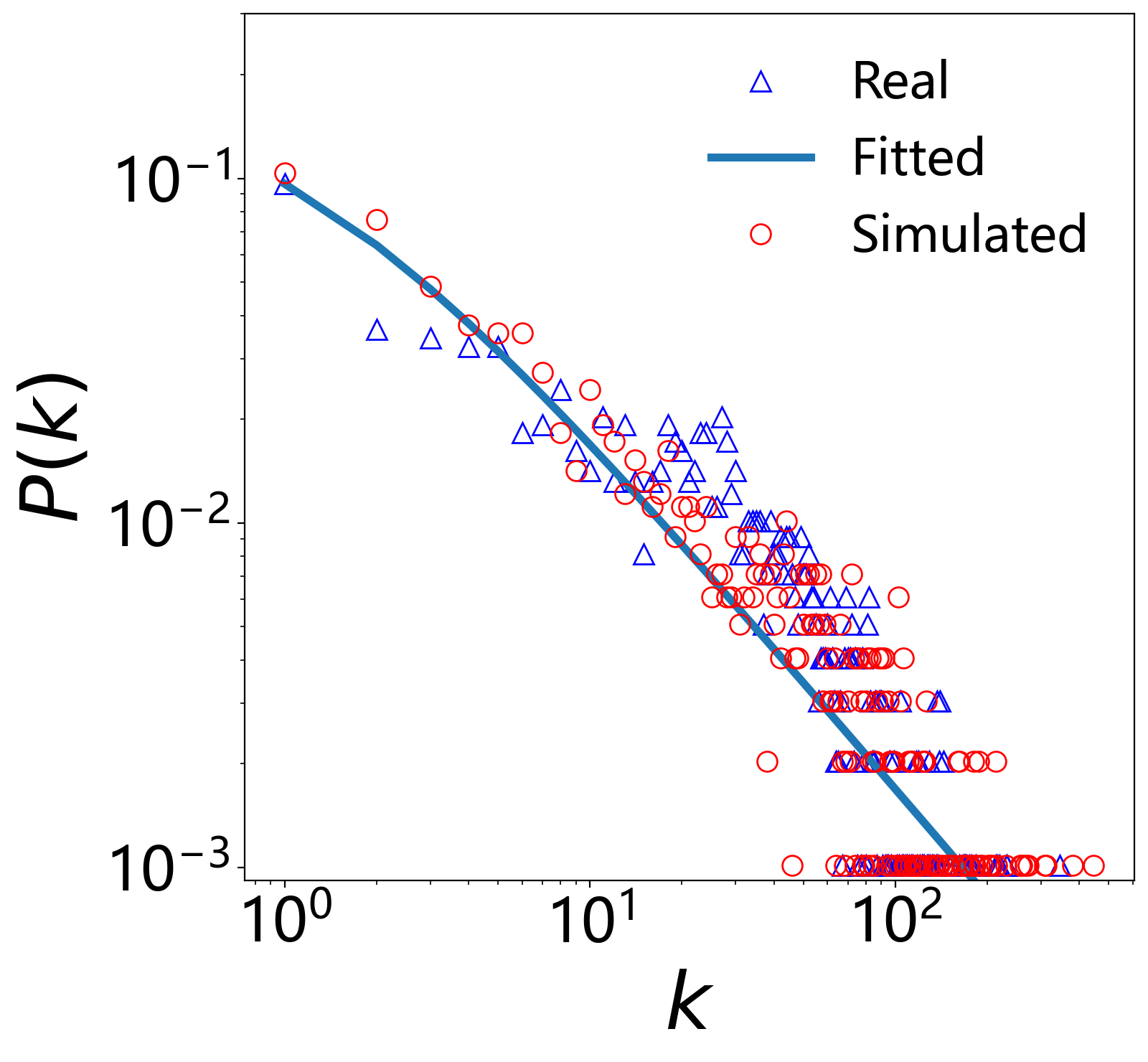}
        \caption{Continuous Time}
        \label{fig5:sub1}
    \end{subfigure}
    \begin{subfigure}[]{0.235\textwidth}
        \includegraphics[width=\textwidth]{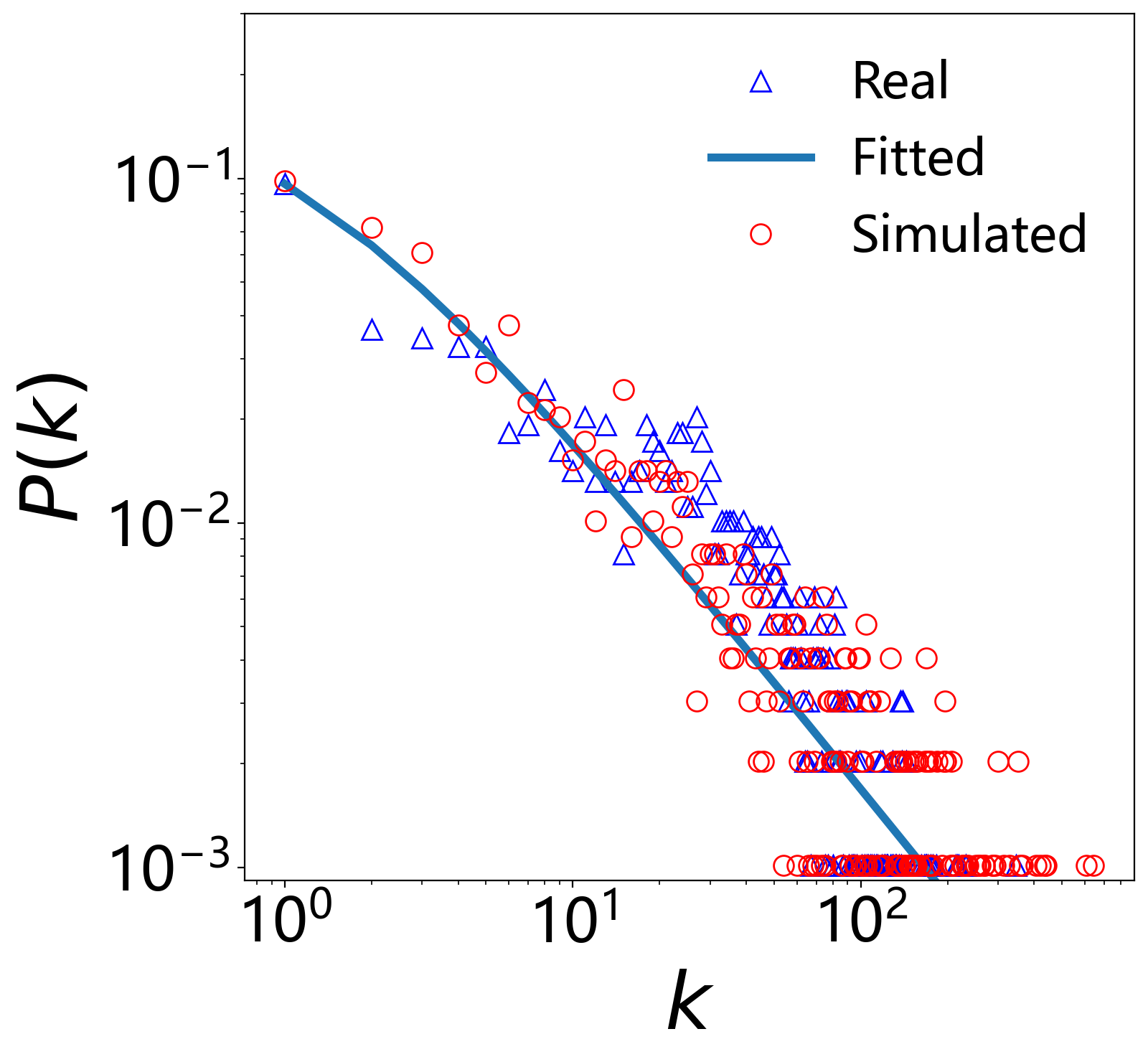}
        \caption{Discrete Time}
        \label{fig5:sub2}
    \end{subfigure}

\caption{\label{fig11}
The fitting results of a continuous and discrete time network model to an Email Network (EN). In these subplots, the blue triangles represent data from the real networks, the blue lines represent the fitting curves, and the red circle denotes the data obtained from the simulation network. }
\end{figure}
\begin{figure}[!ht]
    \begin{subfigure}[]{0.235\textwidth}
        \includegraphics[width=\textwidth]{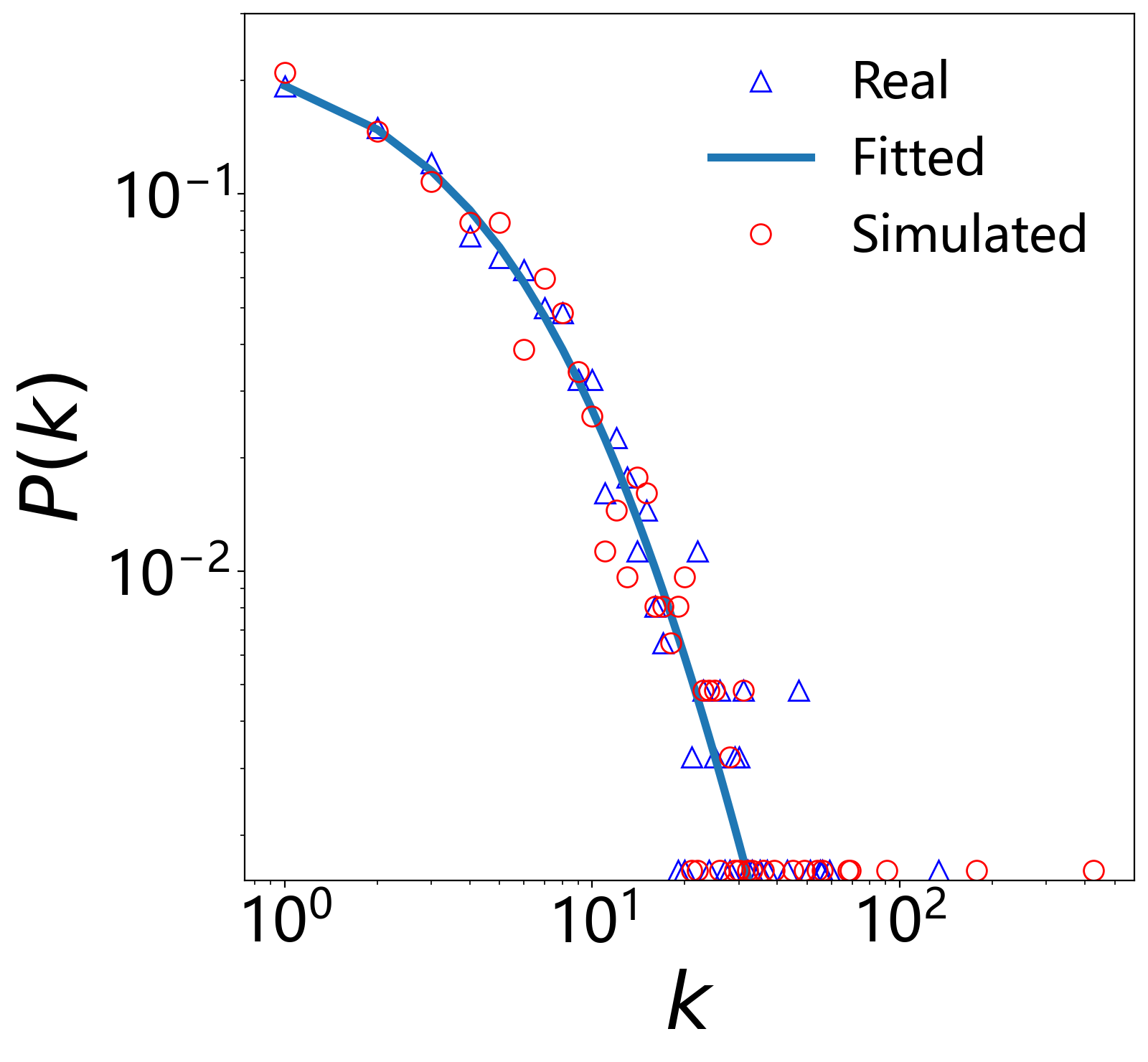}
        \caption{Continuous Time}
        \label{fig5:sub1}
    \end{subfigure}
    \begin{subfigure}[]{0.235\textwidth}
        \includegraphics[width=\textwidth]{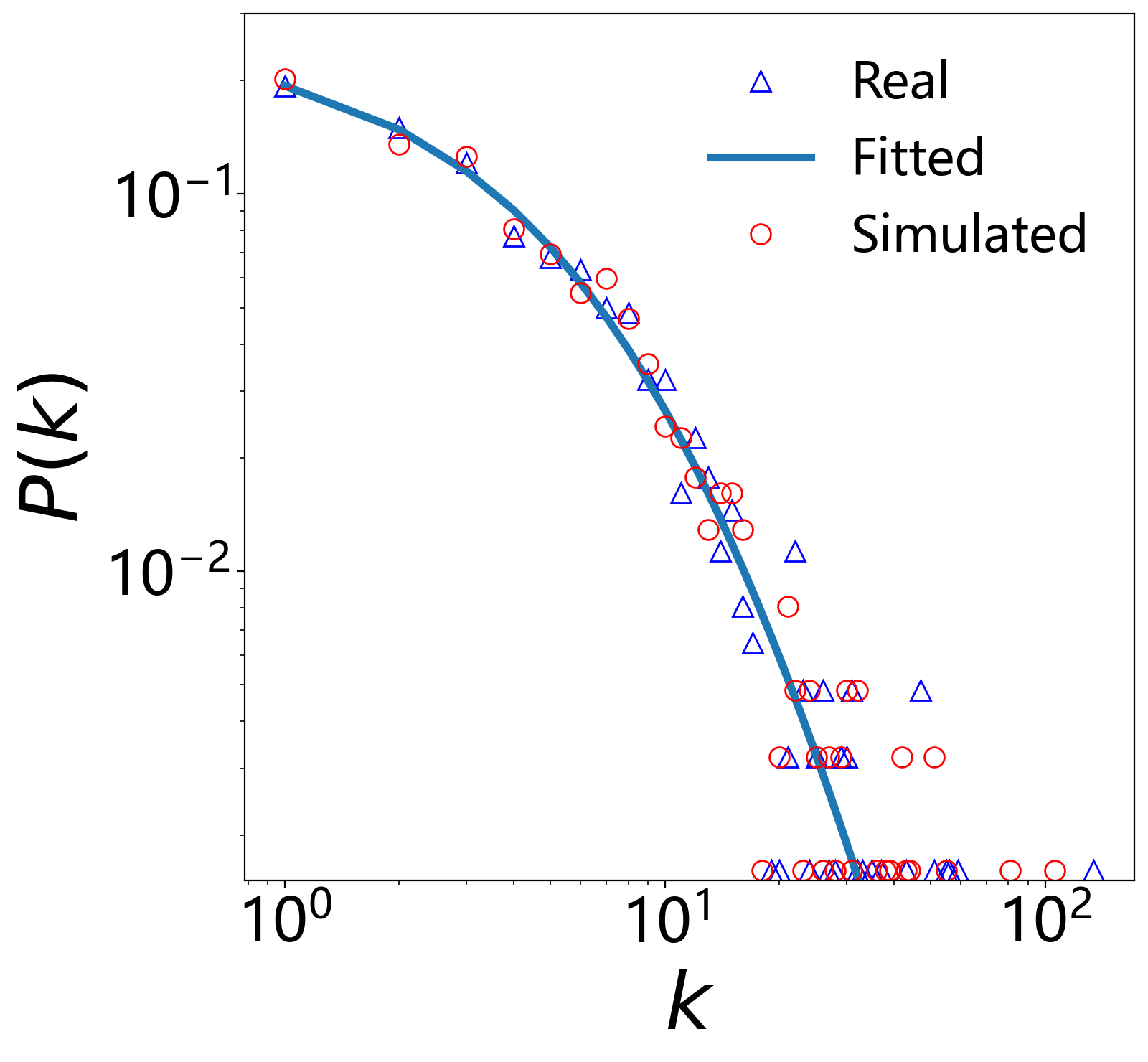}
        \caption{Discrete Time}
        \label{fig5:sub2}
    \end{subfigure}

\caption{\label{fig12}
The fitting results of a continuous and discrete time network model to a Facebook Page network (FPN). In these subplots, the blue triangles represent data from the real networks, the blue lines represent the fitting curves, and the red circle denotes the data obtained from the simulation network. }
\end{figure}

\begin{table*}[h]
\centering
\caption{\label{tab:table3}%
\centering
RESULTS AND PARAMETERS OF DELIBERATE ATTACK THE NETWORKS
}

\begin{tabular}{c|c|c|c|c|c|c|c|c|c}

&
\multicolumn{3}{c|}{FBN}&
\multicolumn{3}{c|}{EN}&
\multicolumn{3}{c}{FPN}\\
\cline{2-10}
&real&continuous&discrete&real&continuous&discrete&real&continuous&discrete\\
\hline
a         & \multicolumn{3}{c|}{2.119}& \multicolumn{3}{c|}{1.040}& \multicolumn{3}{c}{4.586}\\
c         & \multicolumn{3}{c|}{3.382}& \multicolumn{3}{c|}{1.069}& \multicolumn{3}{c}{15.672}\\
\cline{2-10}
KL        &   -   & 0.027& 0.015 &   -    &  0.076 & 0.080 &   -  & 0.040& 0.032   \\
JS        &   -   & \textcolor{black}{0.008}&\textcolor{black}{ 0.004} &   -    &  \textcolor{black}{0.027} &\textcolor{black}{ 0.032} &   -  & \textcolor{black}{0.010}& \textcolor{black}{0.005}   \\
$P_{LCC}$ & 0.994 & 0.995& 0.994 & 1.000  &  1.000 & 1.000 &1.000 &1.000 & 0.995   \\
$<l>$     & 2.910 & 2.267 & 2.874& 2.586  &  2.412 & 2.197 &5.008 &2.603 & 3.971  \\
$C$       & 0.262 & 0.344 & 0.195& 0.407  &  0.290 & 0.543 &0.330 &0.288 & 0.252  \\
\hline
\end{tabular}

\end{table*}

\begin{figure}[!ht]
	\begin{subfigure}[]{0.235\textwidth}
		\hspace{8pt}
		\includegraphics[width=\textwidth]{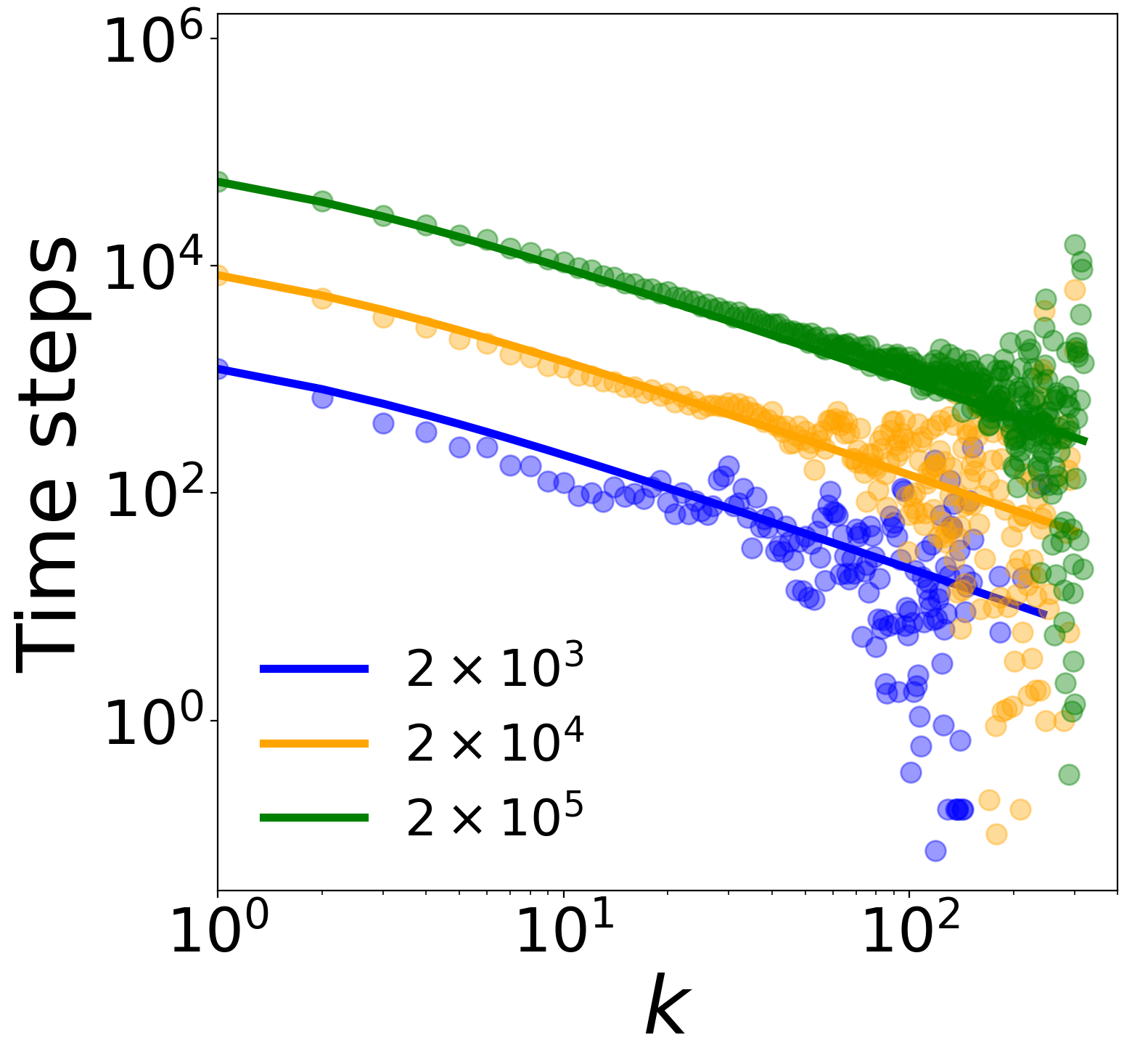}
		\caption{Residence Time}
		\label{fig5:sub1}
	\end{subfigure}
	\begin{subfigure}[]{0.235\textwidth}
		\hspace{8pt}
		\vspace{2.5pt}
		\includegraphics[width=\textwidth]{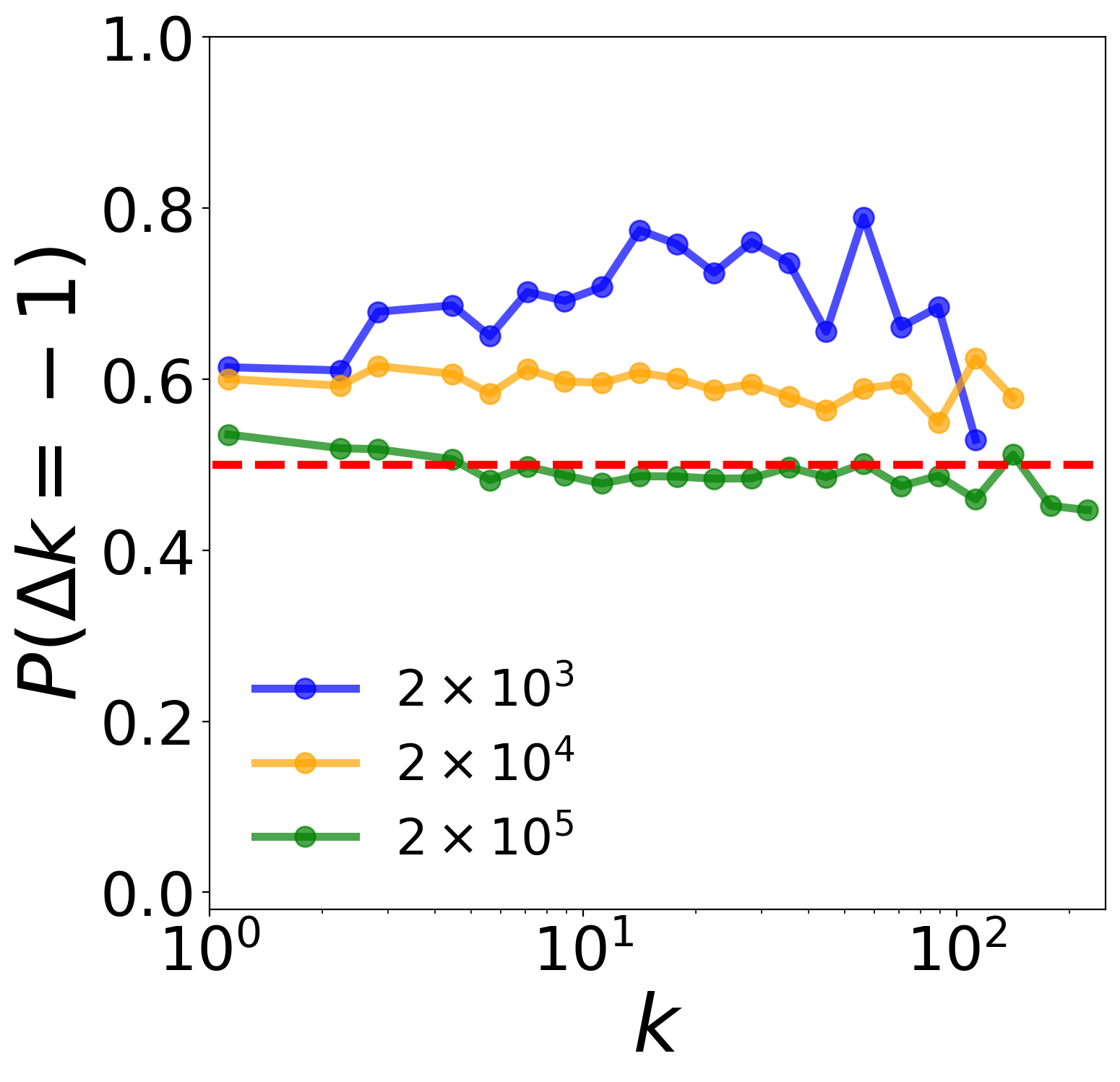}
		\caption{Transition Patterns}
		\label{fig5:sub2}
	\end{subfigure}
	
	\caption{\label{fig13}
		Average residence time and degree transition patterns of nodes in the Email Network (EN). Subplot (a) shows the average residence time of nodes at different degree values, where the scatter points represent the empirical results from the real network and the solid lines denote the simulated results. Subplot (b) shows the probability that nodes with degree $k$ decrease to $k-1$ during a degree transition. Degree values are grouped logarithmically from $10^{0.1x}$ to $10^{0.1(x+1)}$ to reduce fluctuations for large degrees. In both panels, the blue, yellow, and green curves correspond to edge lifetimes of $2 \times 10^3$, $2 \times 10^4$, and $2 \times 10^5$, respectively. The red dashed line in panel b indicates the reference level $P(\Delta k =-1)=0.5$.}
\end{figure}

Figs.~\ref{fig10}-\ref{fig12} and Table~\ref{tab:table3} demonstrate the model's accuracy in replicating the real networks. We utilized a Fly Brain Network (FBN) from networkrepository.com \cite{nr,paranjape2017motifs}, an email-Eu-core temporal network (EN) from snap.stanford.edu, and a mutually liked Facebook pages network about foods (FPN) from networkrepository.com \cite{nr}. The FBN has 1781 nodes and 33641 edges, the EN was generated using email data from a large European research institution. The e-mails only represent communication between institution members, it has 986 nodes and 24929 edges in the static graph. The FPN is about mutually liked Facebook pages, nodes represent the pages and edges are mutual likes among them, it has 620 nodes and 2102 edges.

Initially, the degree distribution of the networks was fitted using Eq.~\ref{eq9} and Eq.~\ref{eq21} to determine the values of $a$ and $c$. Subsequently, these parameters were used to evolve the network model. Figs.~\ref{fig10}-\ref{fig12} show that in the continuous-time network model and discrete network model, the degree distributions of the simulated networks closely match those of the real networks. Table~\ref{tab:table3} presents a comparison of the KL, JS, $P_{LCC}$, average shortest path $<l>$, and average clustering coefficient $C$ between the real and simulated networks. The model aligns well with the connectivity characteristics of all three real-world networks, with the average shortest path length $l$ fitting well for the FBN and EN, though it is smaller for the FPN. The  FPN network has an average shortest path length of 5.008, while the continuous and discrete-time models exhibit shorter path lengths of 2.603 and 3.971, respectively. It is also noticeable that, while the model closely fits $<l>$ for the FBN and EN networks, the evolved networks tend to have generally smaller path lengths.

In terms of clustering coefficient $C$, the model's fit is less accurate compared to $P_{LCC}$ and $<l>$, with larger discrepancies between the model and real networks. For degree distribution, the model fits the real data quite well, with both discrete and continuous-time models showing KL and JS divergences of less than 0.05 for FBN and FPN. However, for EN, the KL and JS divergences are higher, reaching 0.08 and 0.032, respectively. Overall, the proposed network model effectively captures the degree distribution and connectivity features of real scale-free networks, provides a reasonable approximation of the average shortest path length with a slight underestimation, but fails to accurately reproduce clustering characteristics. The relatively high clustering coefficients shown in Table~\ref{tab:table3} can be attributed to the degree-dependent rewiring mechanism under a fixed-size constraint. Since edge creation and deletion events are driven by nodes’ degree-change tendencies, nodes with higher degrees are more likely to interact and form new edges among themselves or with shared neighbors at a higher frequency. This preferential interaction among active high-degree nodes promotes the repeated formation of triangular connections, leading to an elevated level of clustering compared with purely random rewiring models.

To verify the validity of the proposed assumption regarding degree variation patterns in real temporal networks, Fig.~\ref{fig13} presents the residence time and transition behavior of nodes in EN at different degree values. Since the original dataset only records the time when each edge is added but not when it is removed, we assume the edge lifetime to be $2 \times 10^3$, $2 \times 10^4$, and $2 \times 10^5$. As shown in Fig.~\ref{fig13}(a), for all lifetime settings, the residence time exhibits a clear decreasing trend with increasing degree, following an approximately logarithmic decay. This pattern indicates that nodes with higher degrees experience more frequent degree fluctuations. To ensure consistent temporal scaling, the results from the simulated network were rescaled such that the average residence time at $k=1$ matches that of the empirical network. After aligning the temporal scales, the simulated and empirical results remain consistent across all lifetime settings. These findings confirm that the assumption in our model—that nodes with larger degrees are more dynamically unstable—is well supported by the empirical data from EN. Fig.~\ref{fig13}(b) characterizes the directionality of degree transitions by measuring the probability $P(\Delta k=-1)$ that a node’s degree decreases by 1 when a transition occurs. To reduce fluctuations caused by sparse samples at large degrees, degree values are logarithmically binned into intervals of $10^{0.1x}$ - $10^{0.1(x+1)}$, and the average probability within each bin is plotted as a point. As shown in Fig.~\ref{fig13}(b), when the edge lifetime is relatively short, node degrees tend to decrease across all degree ranges, and the current degree value has little influence on the direction of change. When the edge lifetime increases to $2 \times 10^5$ time steps, the probability $P(\Delta k=-1)$ remains consistently close to 0.5 across all degree values. In this case, the observed dynamics are consistent with the modeling assumption in our framework, where nodes are allowed to increase or decrease their degree with equal probability during degree transitions.

\section{Conclusion}

This work introduces a dynamic network model leveraging random walk processes with variable drift coefficients. Stochastic process theory is applied to analyze the stationary degree distribution as the network evolves to a steady state. Depending on the specific setting of the drift coefficients, which is a monotonically increasing function of degree, the network's stationary degree distribution can exhibit a power-law distribution. This enables the modeling of dynamically evolving networks with fixed sizes and power-law degree distributions in both discrete and continuous time. Based on numerical simulations and evaluations using the KL and JS divergences, we find good agreement between the theoretical degree distribution predicted by the model and the simulation results. We also examine the impact of the power-law exponent and attack ratio on the largest connected component of the network and show that, similar to other SF network models, our network model is vulnerable to deliberate attacks but robust against random attacks. In the case of deliberate attacks, we observe the resilience of the largest connected component and the shortest path length. The results indicate that network efficiency takes longer to recover than connectivity. Additionally, the network generated by the discrete-time model exhibits relatively higher robustness. This increased resilience suggests that, under the same attack conditions, the discrete-time model can maintain better connectivity and resilience compared to the continuous-time model. Furthermore, we employ the proposed model to simulate three real-world networks and compared the resulting KL and JS divergences, the size of the largest connected component, the average shortest path length, and the clustering coefficient. Our model fits well with the degree distribution and connectivity of real networks, though it slightly underestimates the average shortest path length and shows suboptimal performance in clustering coefficient fitting. We further validate the rationality of the model’s core assumption using the empirical temporal Email Network. By comparing the degree-dependent residence times and transition probabilities between the real and simulated networks under different edge lifetimes, we observe strong consistency after temporal scale normalization. The results confirm that the modeled mechanism accurately captures the degree fluctuation patterns observed in specific real temporal systems. Such validation is essential for assessing the practical effectiveness and applicability of the proposed model, demonstrating that it can reproduce not only the structural but also the dynamical properties of real-world evolving networks.

The significance of this model lies in its departure from the assumptions of preferential attachment and network growth that underlie conventional SF network models, allowing the generation and simulation of dynamically evolving networks with fixed sizes and controllable power-law exponents in their degree distributions. Additionally, due to the dynamic nature and convergence properties of the network model presented in this work, it can be used to describe and simulate the resilience of specific types of networks after being attacked. This provides a framework for studying the collapse and resilience of network topologies.


%





\ifCLASSOPTIONcaptionsoff
  \newpage
\fi





\bibliographystyle{IEEEtran}
\bibliography{IEEEabrv,Bibliography}

\begin{thebibliography}{10}
\providecommand{\url}[1]{#1}
\csname url@rmstyle\endcsname
\providecommand{\newblock}{\relax}
\providecommand{\bibinfo}[2]{#2}
\providecommand\BIBentrySTDinterwordspacing{\spaceskip=0pt\relax}
\providecommand\BIBentryALTinterwordstretchfactor{4}
\providecommand\BIBentryALTinterwordspacing{\spaceskip=\fontdimen2\font plus
\BIBentryALTinterwordstretchfactor\fontdimen3\font minus
  \fontdimen4\font\relax}
\providecommand\BIBforeignlanguage[2]{{%
\expandafter\ifx\csname l@#1\endcsname\relax
\typeout{** WARNING: IEEEtran.bst: No hyphenation pattern has been}%
\typeout{** loaded for the language `#1'. Using the pattern for}%
\typeout{** the default language instead.}%
\else
\language=\csname l@#1\endcsname
\fi
#2}}

\bibitem{barabasi2004network}
A.-L. Barabasi and Z.~N. Oltvai, ``Network biology: understanding the cell's
  functional organization,'' \emph{Nature reviews genetics}, vol.~5, no.~2, pp.
  101--113, 2004.

\bibitem{wang2009cascade}
J.-W. Wang and L.-L. Rong, ``Cascade-based attack vulnerability on the us power
  grid,'' \emph{Safety science}, vol.~47, no.~10, pp. 1332--1336, 2009.

\bibitem{watts1998collective}
D.~J. Watts and S.~H. Strogatz, ``Collective dynamics of
  ‘small-world’networks,'' \emph{nature}, vol. 393, no. 6684, pp. 440--442,
  1998.

\bibitem{barabasi1999emergence}
A.-L. Barab{\'a}si and R.~Albert, ``Emergence of scaling in random networks,''
  \emph{science}, vol. 286, no. 5439, pp. 509--512, 1999.

\bibitem{albert2000topology}
R.~Albert and A.-L. Barab{\'a}si, ``Topology of evolving networks: local events
  and universality,'' \emph{Physical review letters}, vol.~85, no.~24, p. 5234,
  2000.

\bibitem{bianconi2001bose}
G.~Bianconi and A.-L. Barab{\'a}si, ``Bose-einstein condensation in complex
  networks,'' \emph{Physical review letters}, vol.~86, no.~24, p. 5632, 2001.

\bibitem{li2003local}
X.~Li and G.~Chen, ``A local-world evolving network model,'' \emph{Physica A:
  Statistical Mechanics and its Applications}, vol. 328, no. 1-2, pp. 274--286,
  2003.

\bibitem{li2006modelling}
C.~Li and G.~Chen, ``Modelling of weighted evolving networks with community
  structures,'' \emph{Physica A: Statistical Mechanics and its Applications},
  vol. 370, no.~2, pp. 869--876, 2006.

\bibitem{gu2008local}
Y.~Gu and J.~Sun, ``A local-world node deleting evolving network model,''
  \emph{Physics Letters A}, vol. 372, no.~25, pp. 4564--4568, 2008.

\bibitem{li2017fundamental}
A.~Li, S.~P. Cornelius, Y.-Y. Liu, L.~Wang, and A.-L. Barab{\'a}si, ``The
  fundamental advantages of temporal networks,'' \emph{Science}, vol. 358, no.
  6366, pp. 1042--1046, 2017.

\bibitem{gerlach2019testing}
M.~Gerlach and E.~G. Altmann, ``Testing statistical laws in complex systems,''
  \emph{Physical Review Letters}, vol. 122, no.~16, p. 168301, 2019.

\bibitem{dagum1992dynamic}
P.~Dagum, A.~Galper, and E.~Horvitz, ``Dynamic network models for
  forecasting,'' in \emph{Uncertainty in artificial intelligence}.\hskip 1em
  plus 0.5em minus 0.4em\relax Elsevier, 1992, pp. 41--48.

\bibitem{holme2012temporal}
P.~Holme and J.~Saram{\"a}ki, ``Temporal networks,'' \emph{Physics reports},
  vol. 519, no.~3, pp. 97--125, 2012.

\bibitem{sporns2004organization}
O.~Sporns, D.~R. Chialvo, M.~Kaiser, and C.~C. Hilgetag, ``Organization,
  development and function of complex brain networks,'' \emph{Trends in
  cognitive sciences}, vol.~8, no.~9, pp. 418--425, 2004.

\bibitem{huberman1999growth}
B.~A. Huberman and L.~A. Adamic, ``Growth dynamics of the world-wide web,''
  \emph{Nature}, vol. 401, no. 6749, pp. 131--131, 1999.

\bibitem{bullmore2009complex}
E.~Bullmore and O.~Sporns, ``Complex brain networks: graph theoretical analysis
  of structural and functional systems,'' \emph{Nature reviews neuroscience},
  vol.~10, no.~3, pp. 186--198, 2009.

\bibitem{medo2011temporal}
M.~Medo, G.~Cimini, and S.~Gualdi, ``Temporal effects in the growth of
  networks,'' \emph{Physical review letters}, vol. 107, no.~23, p. 238701,
  2011.

\bibitem{pi2025dynamic}
B.~Pi, L.-J. Deng, M.~Feng, M.~Perc, and J.~Kurths, ``Dynamic evolution of
  complex networks: A reinforcement learning approach applying evolutionary
  games to community structure,'' \emph{IEEE Transactions on Pattern Analysis
  and Machine Intelligence}, vol.~47, no.~10, pp. 8563--8582, 2025.

\bibitem{zhang2016random}
X.~Zhang, Z.~He, and L.~Rayman-Bacchus, ``Random birth-and-death networks,''
  \emph{Journal of Statistical Physics}, vol. 162, pp. 842--854, 2016.

\bibitem{feng2018evolving}
M.~Feng, L.~Deng, and J.~Kurths, ``Evolving networks based on birth and death
  process regarding the scale stationarity,'' \emph{Chaos: An Interdisciplinary
  Journal of Nonlinear Science}, vol.~28, no.~8, 2018.

\bibitem{feng2022heritable}
M.~Feng, Y.~Li, F.~Chen, and J.~Kurths, ``Heritable deleting strategies for
  birth and death evolving networks from a queueing system perspective,''
  \emph{IEEE Transactions on Systems, Man, and Cybernetics: Systems}, vol.~52,
  no.~10, pp. 6662--6673, 2022.

\bibitem{li2023evolving}
Y.~Li, M.~Feng, and J.~Kurths, ``Evolving network modeling driven by the degree
  increase and decrease mechanism,'' \emph{IEEE Transactions on Systems, Man,
  and Cybernetics: Systems}, vol.~53, no.~9, pp. 5369--5380, 2023.

\bibitem{zeng2023temporal}
Z.~Zeng, M.~Feng, and J.~Kurths, ``Temporal network modeling with online and
  hidden vertices based on the birth and death process,'' \emph{Applied
  Mathematical Modelling}, vol. 122, pp. 151--166, 2023.

\bibitem{holme2015modern}
P.~Holme, ``Modern temporal network theory: a colloquium,'' \emph{The European
  Physical Journal B}, vol.~88, pp. 1--30, 2015.

\bibitem{raimundo2018adaptive}
R.~L. Raimundo, P.~R. Guimar{\~a}es, and D.~M. Evans, ``Adaptive networks for
  restoration ecology,'' \emph{Trends in Ecology \& Evolution}, vol.~33, no.~9,
  pp. 664--675, 2018.

\bibitem{petri2018simplicial}
G.~Petri and A.~Barrat, ``Simplicial activity driven model,'' \emph{Physical
  review letters}, vol. 121, no.~22, p. 228301, 2018.

\bibitem{serafino2021true}
M.~Serafino, G.~Cimini, A.~Maritan, A.~Rinaldo, S.~Suweis, J.~R. Banavar, and
  G.~Caldarelli, ``True scale-free networks hidden by finite size effects,''
  \emph{Proceedings of the National Academy of Sciences}, vol. 118, no.~2, p.
  e2013825118, 2021.

\bibitem{katti1968handbook}
S.~K. Katti and A.~V. Rao, ``Handbook of the poisson distribution,''
  \emph{Technometrics}, vol.~10, no.~2, pp. 412--412, 1968.

\bibitem{zhou2020universal}
B.~Zhou, X.~Lu, and P.~Holme, ``Universal evolution patterns of degree
  assortativity in social networks,'' \emph{Social Networks}, vol.~63, pp.
  47--55, 2020.

\bibitem{feng2024information}
M.~Feng, Z.~Zeng, Q.~Li, M.~Perc, and J.~Kurths, ``Information dynamics in
  evolving networks based on the birth-death process: Random drift and natural
  selection perspective,'' \emph{IEEE transactions on systems, man, and
  cybernetics: systems}, 2024.

\bibitem{nr}
\BIBentryALTinterwordspacing
R.~A. Rossi and N.~K. Ahmed, ``The network data repository with interactive
  graph analytics and visualization,'' in \emph{AAAI}, 2015. [Online].
  Available: \url{https://networkrepository.com}
\BIBentrySTDinterwordspacing

\bibitem{paranjape2017motifs}
A.~Paranjape, A.~R. Benson, and J.~Leskovec, ``Motifs in temporal networks,''
  in \emph{Proceedings of the tenth ACM international conference on web search
  and data mining}, 2017, pp. 601--610.

\end{thebibliography}
%

\begin{IEEEbiography}[{\includegraphics[width=1in,height=1.25in,clip,keepaspectratio]{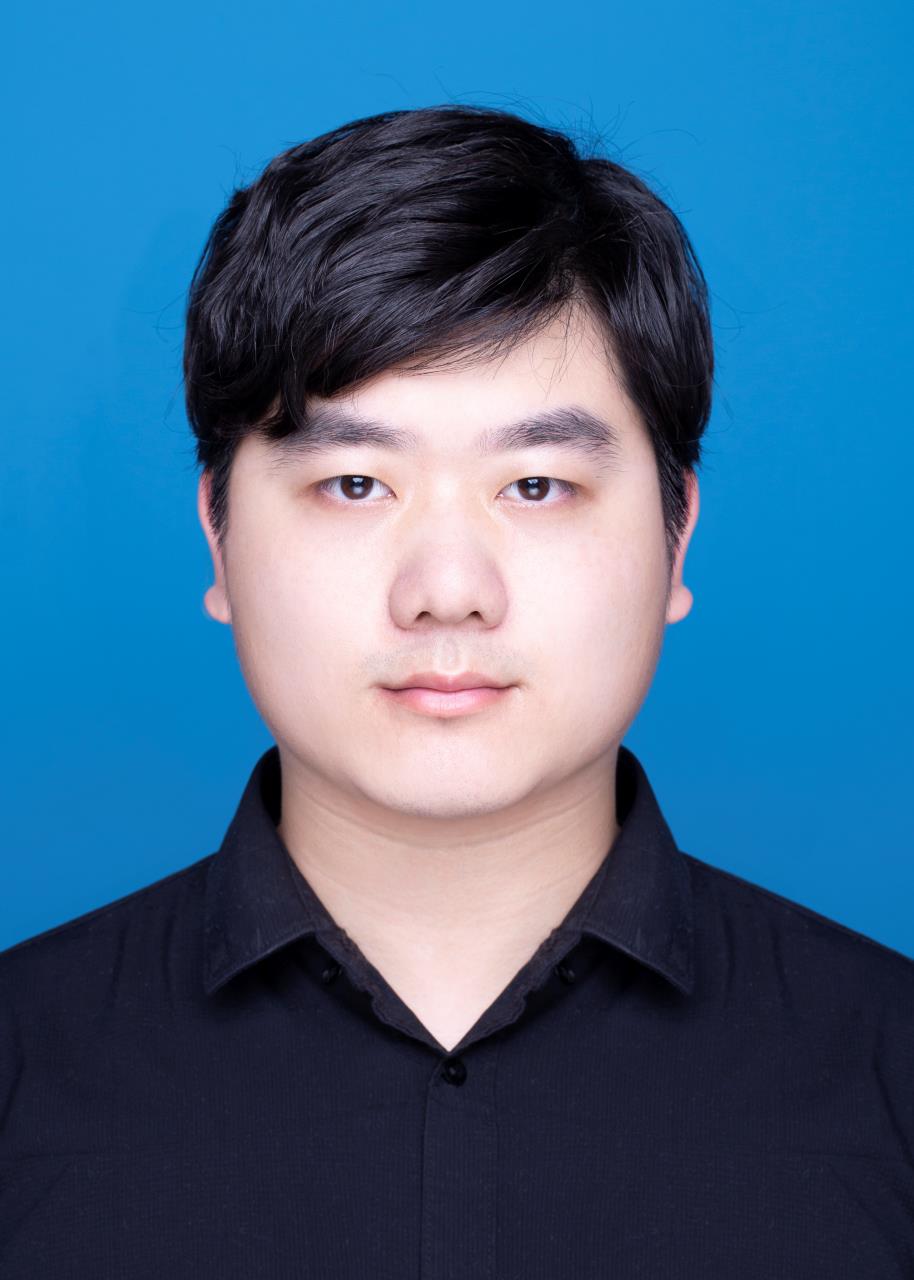}}]{Yichao Yao}
 received the bachelor’s degree from the College of Artificial
Intelligence, Southwest University, Chongqing, China. He is currently pursuing the master’s degree
in computer science. His research interests include complex networks, evolutionary games, stochastic processes, and nonlinear science.
\end{IEEEbiography}

\begin{IEEEbiography}[{\includegraphics[width=1in,height=1.25in,clip,keepaspectratio]{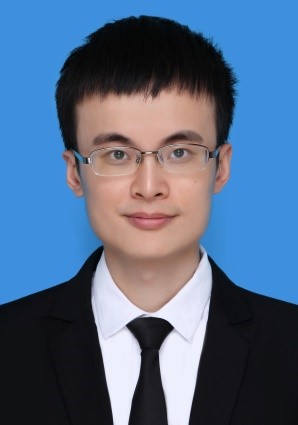}}]{Minyu Feng}
(Senior Member, IEEE) received his Ph.D. degree in Computer Science from a joint program between University of Electronic Science and Technology of China, Chengdu, China, and Humboldt University of Berlin, Berlin, Germany, in 2018. Since 2019, he has been an associate professor at the College of Artificial Intelligence, Southwest University, Chongqing, China. 

Dr. Feng has published more than 70 peer-reviewed papers in authoritative journals, such as IEEE Transactions on Pattern Analysis and Machine Intelligence, IEEE Transactions on Systems, Man, and Cybernetics: Systems, IEEE Transactions on Cybernetics, etc. He is a Senior Member of IEEE, China Computer Federation (CCF), and Chinese Association of Automation (CAA). Currently, he serves as a Subject Editor for Applied Mathematical Modelling, an Editorial Advisory Board Member for Chaos, and an Editorial Board Member for Humanities \& Social Sciences Communications, Scientific Reports, and International Journal of Mathematics for Industry. Besides, he is a Reviewer for Mathematical Reviews of the American Mathematical Society.

Dr. Feng's research interests include Complex Systems, Evolutionary Game Theory, Computational Social Science, and Mathematical Epidemiology.

\end{IEEEbiography}

\begin{IEEEbiography}
[{\includegraphics[width=1in,height=1.25in,clip,keepaspectratio]{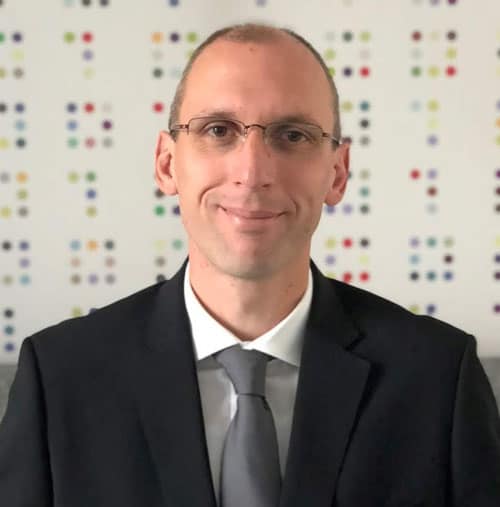}}]{Matja\v{z} Perc} received his Ph.D. in physics from the University of Maribor in 2006. He is currently Professor of physics at the University of Maribor, staff researcher at the Community Healthcare Center Dr. Adolf Drolc Maribor, and Adjunct Professor at Kyung Hee University and Korea University. He is a member of Academia Europaea and the European Academy of Sciences and Arts, and among top 1\% most cited physicists according to 2020, 2021, 2022, 2023, and 2024 Clarivate Analytics data. He is also the 2015 recipient of the Young Scientist Award for Socio and Econophysics from the German Physical Society, and the 2017 USERN Laureate. In 2018 he received the Zois Award, which is the highest national research award in Slovenia. In 2019 he became Fellow of the American Physical Society. Since 2021 he is also Vice-Dean of Natural Sciences at the European Academy of Sciences and Arts.
\end{IEEEbiography}

\begin{IEEEbiography}
[{\includegraphics[width=1in,height=1.25in,clip,keepaspectratio]{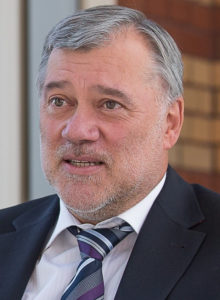}}]{J\"{u}rgen Kurths}
received the B.S. degree in mathematics from the University of Rostock, Rostock, Germany, the Ph.D. degree from the Academy of Sciences, German Democratic Republic, Berlin, Germany, in 1983, the Honorary degree from N.I. Lobachevsky State University, Nizhny Novgorod, Russia in 2008, and the Honorary degree from Saratow State University, Saratov, Russia, in 2012. 

From 1994 to 2008, he was a Full Professor with the University of Potsdam, Potsdam, Germany. Since 2008, he has been a Professor of Nonlinear Dynamics with Humboldt University, Berlin, and the Chair of the Research Domain Complexity Science, Potsdam Institute for Climate Impact Research, Potsdam. He has authored more than 700 papers, which are cited more than 53 000 times (H-index: 106). His main research interests include synchronization, complex networks, time-series analysis, and their applications. 

Dr. Kurths was the recipient of the Alexander von Humboldt Research Award from India, in 2005, and from Poland in 2021, the Richardson Medal of the European Geophysical Union in 2013, and the Eight Honorary Doctorates. He is a Highly Cited Researcher in Engineering. He is an Editor-in-Chief of \textit{CHAOS} and on the editorial boards of more than ten journals. He is a Fellow of the American Physical Society, the Royal Society of Edinburgh, and the Network Science Society. He is a member of the Academia Europaea.
\end{IEEEbiography}





\vfill


\end{document}